\documentclass[aps,superscriptaddress,eqsecnum,nofootinbib,showkeys,preprintnumbers]{revtex4-2}
\usepackage{amsmath}
\usepackage{amssymb}
\usepackage{amsfonts,amsthm,bm}
\usepackage{color,xcolor}
\usepackage[greek,english]{babel}
\usepackage{titlesec}
\usepackage{graphicx,epsfig}
\graphicspath{{figures/}{./}}
\usepackage{float}
\usepackage{tikz}
\usepackage{enumitem}
\usepackage[pdfencoding=auto,colorlinks=true,allcolors=blue]{hyperref}

\newcommand{\n}{\nu}

\newcommand{\abs}[1]{\left| #1 \right|}

\newcommand{\be}{\begin{eqnarray}}
\newcommand{\ee}{\end{eqnarray}}
\newcommand{\bea}{\begin{eqnarray}}

\newcommand{\eea}{\end{eqnarray}}

\newcommand{\beq}{\begin{equation}}
\newcommand{\eeq}{\end{equation}}
\newcommand{\bseq}{\begin{subequations}}
\newcommand{\eseq}{\end{subequations}}

\def\n{\nu}

\def\Ref{\ref}

\begin{document}

\title{Stability of black holes with non-minimally coupled scalar hair to the Einstein tensor}

\author{Nikos Chatzifotis}
\email{chatzifotisn@gmail.com}
\affiliation{Physics Division,
National Technical University of Athens, 15780 Zografou Campus,
Athens, Greece}

\author{Christoforos Vlachos}
\email{cvlach@mail.ntua.gr}
\affiliation{Physics Division,
National Technical University of Athens, 15780 Zografou Campus,
Athens, Greece}

\author{Kyriakos Destounis}
\email{kyriakos.destounis@uni-tuebingen.de}
\affiliation{Theoretical Astrophysics, IAAT, University of T$\ddot{u}$bingen, 72076 T$\ddot{u}$bingen, Germany}

\author{Eleftherios Papantonopoulos}
\email{lpapa@central.ntua.gr}
\affiliation{Physics Division,
National Technical University of Athens, 15780 Zografou Campus,
Athens, Greece}

\vspace{17.5cm}
\begin{abstract}
General relativity admits a plethora of exact compact object solutions. The augmentation of Einstein's action with non-minimal coupling terms leads to modified theories with rich structure, which, in turn, provide non-trivial solutions with intriguing phenomenology. Thus, assessing their viability under generic fluctuations is of utmost importance for gravity theories. We consider static and spherically-symmetric solutions of a Horndeski subclass which includes a massless scalar field non-minimally coupled to the Einstein tensor. Such theory possesses second-order field equations and admits an exact black hole solution with scalar hair. Here, we study the stability of such solution under axial gravitational perturbations and find that it is linearly stable. The qualitative features of the ringdown waveform depend solely on the ratio of the two available parameters of spacetime, namely the black hole mass $m$ and the non-minimal coupling strength $\ell_\eta$. Finally, we demonstrate  the gravitational-wave ringdown transitions between three distinct patterns as the ratio $m/\ell_\eta$ increases; a state which is dominated by photon-sphere excitations and maintains a typical quasinormal ringdown, an intermediate long-lived state which exhibits gravitational-wave echoes and, finally, a state where the ringdown and echoes are depleted rapidly to  turn to an exponential tail.
\end{abstract}

\vspace{3.5cm}
\maketitle
\flushbottom

\section{Introduction}

Compact objects play a decisive role in contemporary astrophysics, as their relativistic collisions may provide crucial information concerning astrophysical processes in extreme-gravity conditions. The latest gravitational-wave (GW) detections by ground-based interferometers \cite{Abbott:2016blz,Abbott:2016nmj,Abbott:2017oio,Abbott:2017vtc,TheLIGOScientific:2017qsa} have provided intelligence regarding the strong-field regime. The early stage of the gravitational ringdown of black hole (BH) mergers, described by quasinormal modes (QNMs) \cite{Vishveshwara:1970zz,Kokkotas:1999bd,Berti:2009kk,Konoplya:2011qq}, further contributes to the understanding of their relaxation properties, as well as the governing theory of gravity. Nevertheless, a conclusive interpretation of the underlying gravitational theory has not yet been met. Therefore, it is expected that future ground and space-borne detectors will improve our perception of gravitational interactions, and in particular will shed light into the existence of exotic compact objects (ECOs) \cite{Mazur:2001fv,Morris:1988tu,Damour:2007ap,Holdom:2016nek,Abedi:2016hgu,Abedi:2020ujo,Destounis:2020kss,Destounis:2021mqv,Destounis:2021rko,Peng:2019cmm} which may possess unexpected multipolar and near-horizon structures that differ significantly from those of BHs.

ECOs are spacetime solutions of general relativity (GR) and modified gravity theories that describe compact objects with exotic properties and intriguing multipolar structure \cite{Barausse:2019pri}, such as BHs which evade the `no-hair' theorem and give rise to additional spacetime parameters (besides the mass, spin and charge), wormholes that evade singularities and connect Universes \cite{Morris:1988tu} as well as horizonless compact objects which possess unexpected near-horizon structures that expel the event horizon and subsequent singularities through reflective centrifugal barriers \cite{Cardoso:2019rvt}. The majority of ECOs, which possess a photon sphere, can naturally mimic BHs when perturbed and produce prompt ringdown waveforms in the time domain which are identical to those of BHs \cite{Cardoso:2016rao,Cardoso:2016oxy}. This occurs due to the indifference of photon sphere excitations from external perturbations. The dominant effects of ECO ringdown only appear at late times in the form of successively damped repetitions of subsequent photon sphere excitation, know as echoes, which occur due to the entrapment of perturbations inside potential wells and the formation of quasibound states \cite{Lasenby:2002mc,Dolan:2007mj,Vieira:2021xqw,Vieira:2021ozg}. These modes represent the actual QNM content of the ECO, which in the frequency domain is dramatically different from the QNMs of BHs \cite{Cheung:2021bol}. In what follows, we will loosely refer to the spectral content of the prompt ringdown as the QNM spectrum whenever the echo timescales are sufficiently large, even though these to do not necessarily correspond to the actual QNMs of the full eigenvalue problem.

Even though GR has withstood many experimental tests, remaining consistent with plenty of observations so far, modified theories of gravity attempt to describe phenomena where GR seems to fail, such as the construction of viable cosmological models for inflation and dark energy \cite{Clifton:2011jh}. The most general scalar-tensor theory of gravity in four dimensions whose Lagrangian is constructed by the metric tensor and a non-minimally coupled scalar field is Horndeski's theory \cite{Horndeski:1974wa}. This gravity theory contains subclasses that preserve a classical Galilean symmetry \cite{Nicolis:2008in,Deffayet:2009wt}, leads to second-order field equations and is free of ghost instabilities \cite{Ostrogradsky:1850fid, Woodard:2006nt, Woodard:2015zca, Deffayet:2011gz}. The most extensively studied subclass of Horndeski theory is represented by a Lagrangian with a scalar field non-minimally coupled to the Einstein tensor.

On large scales, the non-minimal coupling term of the aforementioned subclass possesses very intriguing effects on inflationary dynamics. From an inflationary model-building point of view, it allows for a very effective implementation of a slow-roll phase, due to the fact that it acts as a friction mechanism \cite{Amendola:1993uh,Sushkov:2009hk}, allowing potentials such as the Standard-Model Higgs \cite{Germani:2010hd}, thus making it a very attractive term in Horndeski theory. A generalization of the non-minimal kinetic term and its application to inflation was recently analyzed in \cite{Koutsoumbas:2017fxp,Dalianis:2019vit,Karydas:2021wmx}. During the inflationary phase of the Universe  in GR, scalar and tensor perturbations result in spectra which are red-shifted \cite{DeFelice:2011bh}.  One then expects that the presence of a non-minimal kinetic term will magnify the red-shift behavior of the perturbation spectra because of the friction effect, which is subsequently related to the decrease of the Hubble parameter $H$ during inflation (for a review of the effects of the non-minimal kinetic term in inflation see \cite{Papantonopoulos:2019eff}). On the contrary, if the scalar fields are phantom fields with negative kinetic energy non-minimally coupled to the Einstein tensor, then the spectra of scalar and tensor perturbations produced during an inflationary phase are typically blue-shifted \cite{Baldi:2005gk}. Their dynamics in cosmological setups, as well as the instabilities at which tachyons or ghosts appear in the infrared region around the present Hubble scale were discussed in \cite{Libanov:2007mq}. Beyond inflation, the non-minimal kinetic coupling has also been utilized to construct cosmological models \cite{Saridakis:2010mf}.

Apart from cosmological applications, the particular subclass of Horndeski theory allows the construction of BH solutions with scalar hair \cite{Kolyvaris:2011fk,Rinaldi:2012vy,Kolyvaris:2013zfa,Babichev:2013cya,Charmousis:2014zaa}. Consequently, an important aspect of these compact objects is their stability. Regarding the formation of stable hairy BHs, the `no-hair' theorem should be evaded \cite{Bekenstein:1995un,Hui:2012qt}, which translates to the existence of a balance mechanism to outweigh the gravitational force outside the BH event horizon. A typical example is offered by holography. A charged scalar field theory embedded into an anti-de Sitter (AdS) Lagrangian leads to the formation of horizon hair, as a result of the counterbalance between the attractive gravitational and the repulsive electromagnetic force \cite{Gubser:2008px,Gubser:2005ih}. Then, according to the gauge/gravity duality, such mechanism allows a holographic phase transition which results to a conformal field theory describing a holographic superconductor on the AdS boundary \cite{Hartnoll:2008vx,Hartnoll:2008kx,Kuang:2016edj}, besides other interesting phenomena \cite{Bea:2020ees,Bea:2021zsu,Bea:2021ieq,Bea:2021zol,Bea:2022mfb}.
	
In the subclass of the Horndeski theory in which a scalar field is coupled kinetically to the Einstein tensor there is a direct coupling of matter to curvature and there exist local solutions in which this coupling appears as a primary charge in the metric functions of the resulting hairy BH, that may play the role of an effective negative cosmological constant, even though the action is absent of any cosmological constant term. An interesting aspect of such objects are their thermodynamical properties as well as their viability as novel compact objects. One of the most important requirements for the viability of these objects is their stability against perturbations. To that end, the stability of the hairy BH solution found in \cite{Rinaldi:2012vy}, against linear scalar perturbations, was recently assessed \cite{Vlachos:2021weq}. At the linearized level, the non-minimal coupling constant sources an effective asymptotic boundary where the effective potential of the wave equation that governs the propagation of scalar perturbations diverges. Such a boundary serves as a perfect reflector for incident scalar waves and generates a trapping region outside the photon sphere without the need of invoking a negative cosmological constant in the action of the theory. As a result, the ringdown signal of the BH exhibits successively damped echoes. Thus, scalarized compact objects in the particular Horndeski class can serve as alternatives to the standard echo sources which possess trapping regions beyond the photon sphere due to near-horizon structures \cite{Cardoso:2016rao,Cardoso:2016oxy,Mark:2017dnq,Maselli:2017tfq,Volkel:2018hwb,Konoplya:2018yrp,Cardoso:2019rvt,Maggio:2019zyv,Abedi:2020ujo,Liu:2020qia}.
	
Gravitational perturbations in modified theories of gravity provide information regarding the velocity with which GWs travel. The recent observations of GW170817 and GRB170817A, as well as its electromagnetic counterpart, imply that GWs travel at the speed of light, with deviations smaller than a few $10^{-15}$. The consequences of this experimental result for models of dark energy and modified gravity theories were discussed in  \cite{Baker:2017hug,Creminelli:2017sry}. In particular these constraints on the speed of GWs  were used to test some classes of Horndeski theory. A detailed discussion of the effects of the kinetic coupling on the speed of the GWs in the subclass of the Horndeski theory, in which the scalar field is coupled to the Einstein tensor, is provided in \cite{Gong:2017kim}. It was found that while the kinetic energy of a minimally coupled scalar field does not change under the cosmological evolution, the kinetic energy of the scalar field coupled to the Einstein tensor changes as the Universe expands. At the inflationary epoch it acts as a friction term and drives inflation with steep potentials, while as the Universe expands its contribution to the cosmological evolution is less important and at the late cosmological epoch is negligible, thus GWs propagate at the speed of light at late cosmological times.
	
In any case, Horndeski theories do predict a modified speed of GW propagation. Even so, recent studies \cite{Bahamonde:2019shr,Bahamonde:2019ipm,Bahamonde:2021dqn} demonstrate that with analogue versions of Horndenki gravity, which are based on teleparallel gravity constructed with a nonvanishing torsion tensor, one can device a more general Horndeski theory where GWs propagate with the speed of light without eliminating the coupling functions $G_4(\Phi,X)$ and $G_5(\Phi,X)$ that were highly constrained in standard Horndeski theory. Hence, in the teleparallel approach one is able to restore these terms, creating an interesting way to revive this theory of gravity. Even though our analysis still lies in a curvature-based formulation of gravity, it is still very interesting that there are ways of evading the tight constraints of Horndeski theory.
	
The purpose of this work is twofold. First, we investigate the effect of the kinetic coupling of the scalar field to the Einstein tensor on the stability of local solutions of the particular subclass of Horndeski theories. We will work with the hairy BH solution \cite{Rinaldi:2012vy} for which scalar perturbations have been analyzed recently \cite{Vlachos:2021weq,Chatzifotis:2020oqr} for a wide range of the kinetic coupling. Under these analyses the hairy BH was found to be linearly stable with echoes being present at late times on the ringdown waveform. Here, we perform a first step towards gravitational modal stability, thus extending our previous test scalar field analysis to axial gravitational perturbations. Since the kinetic coupling appears as a primary charge in the metric functions, we expect to get a better understanding on the stability of such objects, although a complete picture of gravitational modal stability can only be discerned when one considers not only the axial but also the polar sector of fluctuations which generally couple to the scalar hair present in  scalar-tensor theories. Our second goal is to investigate how the kinetic coupling affects the ringdown waveform and attempt to assess the appearance of GW echoes in the parametric space of such geometries.

The work is organized as follows. In Section \ref{sec2} we review the BH solution of the Horndeski theory with a scalar field kinetically coupled to the Einstein tensor. In Section \ref{sec:linear axial} we discuss the general framework of axial gravitational perturbations and we derive the effective potential of the considered BH solution. In Section \ref{sec4} we demonstrate the numerical scheme of time-domain integration. In Section \ref{sec5} we study the evolution of the axial gravitational perturbations and finally in Section \ref{sec6} we conclude this work.

\section{Black hole solution with a scalar field kinetically coupled to Einstein tensor}
\label{sec2}

In what follows, we will consider static solutions of a scalar-tensor theory in which the scalar field is kinetically coupled to the Einstein tensor. This is part of the most general scalar-tensor theory which yields second-order field equations, namely the Horndeski theory. The full Lagrangian is given by:
\begin{align}
	\label{1.1}
	\mathcal{L}&=\sum_{i=2}^{i=5}\mathcal{L}_i~,\\
	\nonumber
	\mathcal{L}_2&=K(\Phi,X)~,\\
	\nonumber
	\mathcal{L}_3&=-G_3(\Phi,X)\square\Phi~,\\
	\nonumber
	\mathcal{L}_4&=G_4(\Phi,X)R+G_{4,X}\left[(\square\Phi)^2-(\nabla_\mu\nabla_\nu\Phi)^2\right]~.\\
	\nonumber
	\mathcal{L}_5&=G_5(\Phi,X)G_{\mu\nu}\nabla^\mu\nabla^\nu\Phi-\frac{1}{6}G_{5,X}\left[(\square\Phi)^3-3\square\Phi(\nabla_\mu\nabla_\nu\Phi)^2
+2(\nabla_\mu\nabla_\nu\Phi)^3\right]~,
\end{align}
where $X=-\frac{1}{2}\nabla_\mu\Phi\nabla^\mu\Phi$. We consider a particular subset of Horndeski theory with non-trivial $\mathcal{L}_2=K(\Phi,X)=2\varepsilon X$ and $G_4(\Phi,X)=(8\pi)^{-1}-\eta X$ terms. The theory is described by the following action,
\begin{equation}
	\label{1.2}
	S=\int d^4x \sqrt{-g}\left[\frac{R}{8\pi}-(\varepsilon g_{\mu\nu}+\eta G_{\mu\nu})\partial^{\mu}\Phi\partial^{\nu}\Phi\right],
\end{equation}
where $g_{\mu\nu}$ is the metric tensor, $g={\rm det}(g_{\mu\nu})$, $R$ is the scalar curvature, $G_{\mu\nu}$ is the Einstein tensor, $\Phi$ is a real massless scalar field and $\eta$ is the non-minimal kinetic coupling parameter with dimensions of length-squared. The $\varepsilon$ parameter equals $\pm 1$, where in the case $\varepsilon=+1$ we have a canonical scalar field with positive kinetic term, while the case $\varepsilon=-1$ corresponds to a phantom scalar field with negative kinetic energy. Even though in the original Hordenski theory the kinetic energy of the scalar field is positive, in this work we will also considered the case where the scalar field's kinetic energy is negative.

Varying the action (\Ref{1.2}) with respect to the metric tensor $g_{\mu\nu}$ and scalar field $\Phi$ provides the following field equations
\begin{subequations}
	\label{1.3}
	\begin{align}
		\label{1.3a}
		& G_{\mu\nu}=8\pi\big[\varepsilon T_{\mu\nu}
		+\eta \Theta_{\mu\nu}\big]~, \\
		\label{1.3b}
		&[\varepsilon g^{\mu\nu}+\eta G^{\mu\nu}]\nabla_{\mu}\nabla_\nu\Phi=0~,
	\end{align}
\end{subequations}
where
\begin{align}
	\label{1.4}
	T_{\mu\nu}&=\nabla_\mu\phi\nabla_\nu\Phi-
	\frac{1}{2}g_{\mu\nu}(\nabla\Phi)^2~, \\
	\Theta_{\mu\nu}&=-\frac{1}{2}R\nabla_\mu\Phi\,\nabla_\nu\Phi\,
	+2\nabla_\alpha\Phi\,\nabla_{(\mu}\Phi R^\alpha_{\nu)}
	+\nabla^\alpha\Phi\,\nabla^\beta\Phi\,R_{\mu\alpha\nu\beta}
	+\nabla_\mu\nabla^\alpha\Phi\,\nabla_\nu\nabla_\alpha\Phi
	\nonumber\\
	&-\nabla_\mu\nabla_\nu\Phi\,\square\Phi-\frac{1}{2}(\nabla\Phi)^2
	G_{\mu\nu}
	+g_{\mu\nu}\big[-\frac{1}{2}\nabla^\alpha\nabla^\beta\Phi\,
	\nabla_\alpha\nabla_\beta\Phi
	+\frac{1}{2}(\square\Phi)^2 -\nabla_\alpha\Phi\,\nabla_\beta\Phi\,R^{\alpha\beta}
	\big]~. \label{1.5}
\end{align}

A static and spherically-symmetric BH solution to the aforementioned theory has been found in \cite{Rinaldi:2012vy}, where the scalar field of the theory depends only on the radial coordinate. The solution yields the constraint $\varepsilon\eta<0$, which leads to the definition of the following coupling parameter
\begin{equation}
	\label{2.1}
	\ell_\eta=\abs{\varepsilon\eta}^{1/2}~.
\end{equation}
In terms of the line element
\begin{equation}
	\label{2.2}
	ds^2=-g_{tt}(r)dt^2+g_{rr}(r)dr^2+g_{\theta\theta}(r)d\Omega^2~,
\end{equation}
the BH solution corresponds to $g_{\theta\theta}(r)=r^2$ with $r\in(0,+\infty)$ and yields the following metric components
\begin{subequations}
	\label{2.3}
	\begin{align}
		\label{2.3a}
		g_{tt}(r)&=-\frac{1}{4}F(r)~,\\
		\label{2.3b}
		g_{rr}(r)&=\frac{(r^2+2\ell_\eta^2)^2}{(r^2+\ell_\eta^2)^2F(r)}~,\\
		F(r)&=\left[3+\frac{r^2}{3 \ell^2_\eta}-\frac{8 m}{r}+\frac{\ell_\eta}{r}\arctan\left(\frac{r}{\ell_\eta}\right)\right],
	\end{align}
\end{subequations}
while the scalar hair of the theory reads
\begin{equation}
	\label{2.4}
	(\partial_r\Phi)^2=\Psi^2=-\frac{\varepsilon}{8\pi \ell_\eta^2}\frac{r^2 g_{rr}}{r^2+\ell_\eta^2}~,
\end{equation}
which implies that
\begin{equation}
	\label{2.5}
	4\pi\eta\Psi^2=\frac{1}{2}\frac{r^2 }{r^2+\ell_\eta^2}g_{rr}> 0~, \qquad \forall r> r_h
\end{equation}
where $r=r_h$ is the event horizon radius. It is important to note that the asymptotic behavior of the lapse function \eqref{2.3a} when $r\rightarrow+\infty$, becomes $F(r)\sim r^2/\ell_\eta^2$, where the term $\ell_\eta^2$ assumes a form of an effective cosmological scale, similar to that of an actual cosmological radius, with dimensionality length squared in geometrized units. Even so, we have to stress that the action does not contain any negative cosmological constant term and the emergence of this effective scale is solely due to the non-minimal coupling of the scalar field to the Einstein tensor.

Another important note is that the equations of motion of the scalar field, (\Ref{1.3b}), can be expressed as the conservation of the Noether current that corresponds to the shift symmetry of the Galileon, i.e. $\Phi\rightarrow\Phi+\delta\Phi$, where $\delta\Phi$ is constant. It can straightforwardly be found that the current is defined as
\begin{equation}
	\label{2.6}
	J^\mu=(\varepsilon g^{\mu\nu}+\eta G^{\mu\nu})\nabla_\nu\Phi~.
\end{equation}	
The BH solution satisfies the physical requirement that the norm of this current does not diverge at the horizon, by virtue of (\Ref{2.1}). The scalar hair, however, diverges at the horizon as one can readily see from (\Ref{2.4}). One can also deduce from (\Ref{2.4}) that the metric components can be expressed in terms of the scalar hair. As such, the scalar hair of the theory can be understood as an intrinsic part of the geometry. Finally, we note that due to the Bianchi identity, $\nabla^\mu G_{\mu\nu}=0$, the equation (\Ref{1.3a}) leads to a differential consequence
\begin{equation}
	\label{2.7}
	\nabla^\mu\big[\varepsilon T_{\mu\nu}+\eta \Theta_{\mu\nu}\big]=0~.
\end{equation}
The substitution of expressions (\Ref{1.4}) and (\Ref{1.5}) into the Bianchi identity yields Eq. (\Ref{1.3b}). In other words, the equation of motion of scalar fields (\Ref{1.3b}) is the differential consequence of the system (\Ref{1.3a}) due to the general covariance and the absence of further degrees of freedom. Let us also note that the solution reproduces the Schwarzschild BH in the limit of $\ell_\eta\rightarrow+\infty$, therefore the metric can be understood as a hairy BH generalization of the Schwarzschild spacetime with effective AdS-asymptotics, when the spin-0 degree of freedom also acquires dynamics from the kinetic mixing with the graviton, i.e. the $G^{\mu\nu}\partial_\mu\Phi\partial_\nu\Phi$ term.

As one can observe, the metric functions (\ref{2.3}) of the BH solution \cite{Rinaldi:2012vy} depend only on the parameter $\ell_\eta$, besides the mass parameter $m$. Therefore, they do not contain the information of whether the produced compact object is made of normal or phantom matter as long as $\epsilon\eta<0$. Even though in \cite{Rinaldi:2012vy} the scalar field is assumed to be canonical ($\epsilon>0$) with the non-minimal coupling being negative ($\eta<0$), we have checked that the same solution is obtained when the scalar field is phantom ($\epsilon<0$) and the non-minimal coupling is positive ($\eta>0$).

In finality, let us mention that in \cite{Sushkov:2009hk} it was argued that the subclass of the Horndeski theory under consideration yields wormhole solutions as well. We wish to note here that the wormhole solution derived there is just a coordinate artifact of the BH and does not correspond to a traversable wormhole. Let us consider the BH metric we previously described
\begin{equation}
	ds^2=-\frac{1}{4}F(r)dt^2+\frac{(r^2+2\ell_\eta^2)^2}{(r^2+\ell_\eta^2)^2F(r)}dr^2+r^2d\Omega^2.
\end{equation}
We perform the following coordinate transformation
\begin{equation}\label{coord_trans}
	u^2=r^2-a^2, \qquad u\in  (-\infty,+\infty)
\end{equation}
and the coordinate redefinition
\begin{equation}
	\label{dt}
	dt^2=\frac{4}{\left(3-\frac{8m}{a}+\frac{a^2}{3\ell_\eta^2}+\frac{\ell_\eta}{a}\arctan\left(\frac{a}{\ell_\eta}\right)\right)}dT^2=CdT^2\,.
\end{equation}
Note that this coordinate transformation covers the BH region for $r>a$ twice. Fixing $a>r_h$, the corresponding geometry will cover only the region $r>r_h$ twice. Performing the coordinate transformation \eqref{coord_trans}, one finds:
\begin{align}
	\nonumber
	ds^2&=-CF(\sqrt{u^2+a^2})dT^2\\
	\nonumber
	&+\frac{u^2(u^2+a^2+2\ell_\eta^2)^2}{(u^2+a^2)(u^2+a^2+\ell_\eta^2)^2F(\sqrt{u^2+a^2)}}du^2\\
	\label{2.16}
	&+(u^2+a^2)d\Omega^2,\\
	F(\sqrt{u^2+a^2)}&=\left(3-\frac{8m}{\sqrt{u^2+a^2}}+\frac{u^2+a^2}{3\ell_\eta^2}+\frac{\ell_\eta}{\sqrt{u^2+a^2}}\arctan\left(\frac{\sqrt{u^2+a^2}}{\ell_\eta}\right)\right).
\end{align}
The corresponding metric is the solution derived in \cite{Korolev:2014hwa}, modulo the form of the $g_{tt}$ component, which was left as an indefinite integral. In particular, the integral factor in the $g_{tt}$ component found in \cite{Korolev:2014hwa} can be solved exactly to yield:
\begin{align}
	\nonumber
	g_{tt} &= \frac{a}{\sqrt{u^2+a^2}}\exp\left[\int_0^u\frac{u(u^2+a^2+2\ell_\eta^2)^2} {\ell_\eta^2(u^2+a^2)(u^2+a^2+\ell_\eta^2)F(\sqrt{u^2+a^2})}dr\right]\\
	\nonumber
	&=\frac{a}{\sqrt{u^2+a^2}}\exp\left[\int_0^u\left(\frac{d(\ln[\sqrt{u^2+a^2}F(\sqrt{u^2+a^2})])}{du}\right)du\right]\\
	&=\frac{F(\sqrt{u^2+a^2})}{F(a)}\overset{(\Ref{dt})}{=} \frac{C}{4}F(\sqrt{u^2+a^2})
\end{align} Obviously, a compact object cannot change nature due to a coordinate transformation, thus, the metric (\Ref{2.16}) is just the BH solution \eqref{2.3} written in a ``bad'' coordinate system and was falsely identified as a wormhole. This result is in accordance with the findings in \cite{Evseev:2017jek} regarding the absence of static and spherically-symmetric wormhole solutions in the particular subclass of Horndeski theory.

\section{Axial gravitational perturbations: general analysis}\label{sec:linear axial}

In this section we will undergo an analysis of axial perturbations of the BH solution using Chandrasekhar's method \cite{Chandrasekhar}. The most general metric for an axisymmetric non-stationary spacetime is given by
\begin{equation}
	\label{3.1}
	ds^2=-e^{2f_0}dt^2+e^{2f_3}(d\phi-q_0dt-q_1 dr-q_2 d\theta)^2+e^{2f_1}dr^2+e^{2f_2}d\theta^2~.
\end{equation}
This result is found by use of the Cotton-Darboux theorem, which states that any three-dimensional metric, $g=g^{ij}\partial_i\partial_j$, can always be brought to a diagonal form by a local coordinate transformation. It is clear that the background metric of our solutions can be described by $q_i=0$. In this gauge, axial perturbations are described by the non-vanishing values of $q_i$, while polar perturbations are described by $f_i\rightarrow f_i+\delta f_i$ and $q_i=0$.

For the purposes of writing down the explicit form of the equations (\Ref{1.3a}) for the most general form of the metric of (\Ref{3.1}), we shall obtain the components of the curvature tensors via Cartan's structure equations. We choose the following tetrads to work with
\begin{subequations}
	\begin{align}
		\label{3.2a}
		&\epsilon^{0}_{\mu}=(e^{f_0},0,0,0)~,\\
		\label{3.2b}
		&\epsilon^{1}_{\mu}=(0,e^{f_1},0,0)~,\\
		\label{3.2c}
		&\epsilon^{2}_{\mu}=(0,0,e^{f_2},0)~,\\
		\label{3.2d}
		&\epsilon^{3}_{\mu}=(-e^{f_3}q_0,-e^{f_3}q_1,-e^{f_3}q_2,e^{f_3})~.
	\end{align}
\end{subequations}
Under these tetrads, the basis is found to be
\begin{subequations}
	\begin{align}
		\label{3.3a}
		\epsilon^{0}=e^{f_0}dt&\rightarrow dt=e^{-f_0}\epsilon^0~,\\
		\label{3.3b}
		\epsilon^{1}=e^{f_1}dt&\rightarrow dr=e^{-f_1}\epsilon^1~,\\
		\label{3.3c}
		\epsilon^{2}=e^{f_2}d\theta&\rightarrow d\theta=e^{-f_2}\epsilon^2~,\\
		\label{3.3d}
		\epsilon^{3}=-e^{f_3}q_0dt-e^{f_3}q_1dr-e^{f_3}q_2d\theta+e^{f_3}d\phi&\rightarrow d\phi=e^{-f_3}\epsilon^3+q_0e^{-f_0}\epsilon^0+q_1e^{-f_1}\epsilon^1+q_2e^{-f_2}\epsilon^2~.
	\end{align}
\end{subequations}
The reasoning behind these tetrads is that they associate the perturbations to a single tetrad and thus allow for a decoupling of the equations of motion at first order. The spin connections can be derived from the tetrad postulate, where we associate the zero torsion condition with the Levi-Civita connection as
\begin{equation}
	\label{3.4}
	\omega^a_{\mu b}=\epsilon^a_{\nu}\epsilon^{\lambda}_b\Gamma^{\nu}_{\mu\lambda}-\epsilon^{\lambda}_b\partial_{\mu}\epsilon^a_{\lambda}~.
\end{equation}
Using Cartan's second structure equation, we can derive the Riemann tensor,
\begin{equation}
	\label{3.5}
	R^{a}_{b\mu\nu}=2(\partial_{[\mu}\omega^a_{\n] b}+\omega^{a}_{c[\mu }\omega^c_{\nu] b})~,
\end{equation}
and from (\Ref{3.5}) all the necessary tensors for the equations of motion of the underlying gravity theory. From (\Ref{1.3a}), we know that
\begin{equation}
	\label{3.6}
	G_{ab}=\hat{T}_{ab}~,	
\end{equation}
where $\hat{T}_{ab}=8\pi\big[\varepsilon T_{ab}+\eta \Theta_{ab}\big]$ (note that indices $a,b$ are Lorentz and not spacetime indices). However, the Einstein and stress-energy tensors acquire contributions from the perturbations. From the linearization of the equations of motion we find that only the $G_{03},G_{13},G_{23}$ terms are important at first order. In fact, equation $\delta G_{03}=\delta \hat{T}_{03}$ is degenerate, i.e. it is automatically satisfied by the other two equations. In particular, we find the following results
\begin{subequations}
	\label{3.7}
	\begin{align}
		\label{3.7a}
		&\delta G_{13}=\frac{1}{2r^3\sin^2\theta}\frac{1}{\sqrt{g_{tt}}}\frac{\partial}{\partial\theta}Q+\frac{r\sin\theta}{2g_{tt}\sqrt{grr}}\frac{\partial}{\partial t}\left[\frac{\partial}{\partial r}q_a
-\frac{\partial}{\partial t}q_b\right]~,\\
		\nonumber \\
		\label{3.7b}
		&\delta \hat{T}_{13}=\frac{4\pi\eta\Psi^2}{g_{rr}}\delta G_{13}~,\\
		\nonumber \\
		\label{3.7c}
		&\delta G_{23}=-\frac{1}{2r^2\sin^2\theta}\frac{1}{\sqrt{g_{tt}g_{rr}}}\frac{\partial}{\partial r}Q+\frac{\sin\theta}{2g_{tt}}\frac{\partial}{\partial t}\left[\frac{\partial}{\partial\theta}q_a
-\frac{\partial}{\partial t}q_c\right]~,\\
		\nonumber \\
		\label{3.7d}
		&\delta \hat{T}_{23}=\frac{4\pi\eta\Psi^2}{g_{rr}}\left(\frac{1}{2r^2\sin^2\theta}\frac{-1}{\sqrt{g_{tt}g_{rr}}}\frac{\partial}{\partial r}Q
-\frac{\sin\theta}{2g_{tt}}\frac{\partial}{\partial t}\left[\frac{\partial}{\partial\theta}q_a-\frac{\partial}{\partial t}q_c\right]\right)\\
		\nonumber \\
		&
		\nonumber
		\qquad	-4\pi\eta\frac{\partial}{\partial r}\left(\frac{\Psi^2}{2g_{rr}}\right)\frac{1}{r^2\sin^2\theta \sqrt{g_{tt}}\sqrt{g_{rr}}}Q~,
	\end{align}
\end{subequations}
where $\displaystyle{Q=\left[r^2\sin^3\theta\frac{\sqrt{g_{tt}}}{\sqrt{g_{rr}}}\frac{\partial}{\partial\theta}q_b-r^2\sin^3\theta\frac{\sqrt{g_{tt}}}{\sqrt{g_{rr}}}\frac{\partial}{\partial r}q_c\right]}$. Using the redefinition
\begin{equation}
	\label{3.8}
	\mathcal{F}=\left(1-\frac{4\pi\eta\Psi^2}{grr}\right)Q~,
\end{equation} the differential equations we get from $\delta G_{ab}=\delta \hat{T}_{ab}$, using (\Ref{3.7a}-\Ref{3.7d}), read
\begin{subequations}
	\label{3.9}
	\begin{align}
		\label{3.9a}
		\frac{1}{r^4\sin^3\theta}\sqrt{g_{tt}}\sqrt{g_{rr}}\left(\frac{g_{rr}}{grr-4\pi\eta\Psi^2}\right)\frac{\partial}{\partial\theta}\mathcal{F}&=\frac{\partial}{\partial t}\left[\frac{\partial}{\partial t}q_b
-\frac{\partial}{\partial r}q_a\right]~,\\
		\nonumber\\
		\label{3.9b}
		\frac{1}{r^2\sin^3\theta}\frac{\sqrt{g_{tt}}}{\sqrt{g_{rr}}}\left(\frac{g_{rr}}{grr+4\pi\eta\Psi^2}\right)\frac{\partial}{\partial r}\mathcal{F}&=\frac{\partial}{\partial t}\left[\frac{\partial}{\partial\theta}q_a
-\frac{\partial}{\partial t}q_c\right]~.
	\end{align}
\end{subequations}
Differentiating (\Ref{3.9a}) and (\Ref{3.9b}) by $\theta$ and $r$ respectively, and adding them together yields the differential equation that governs axial perturbations
\begin{equation}
	\label{3.10}
	\frac{r^4}{\sqrt{g_{tt}}\sqrt{g_{rr}}}\left(\frac{g_{rr}
-4\pi\eta\Psi^2}{g_{rr}}\right)\frac{\partial}{\partial r}\left[\frac{1}{r^2}\frac{\sqrt{g_{tt}}}{\sqrt{g_{rr}}}\left(\frac{g_{rr}}{grr
+4\pi\eta\Psi^2}\right)\frac{\partial}{\partial r}\mathcal{F}\right]-\frac{r^2}{g_{tt}}\frac{\partial^2\mathcal{F}}{\partial t^2}
+\sin^3\theta\frac{\partial}{\partial\theta}\left[\frac{1}{\sin^3\theta}\frac{\partial}{\partial\theta}\mathcal{F}\right]=0~.
\end{equation}
Using a separation of variables, the angular component of equation (\Ref{3.10}) can be understood as the known ultraspherical differential equation with solutions the Gegenbauer polynomials, i.e. the angular component of the perturbation is the same as in the Schwarzschild case. This is to be expected, since both spacetimes are spherically symmetric. Thus, setting $\mathcal{F}=\mathcal{R}(r,t)\mathcal{S}(\theta)$, where in the Appendix \ref{appA} we explicitly show that $\mathcal{S}(\theta)=C^{-3/2}_{\ell+2}(\theta)$, (\Ref{3.10}) can be rewritten as
\begin{equation}
	\label{3.11}
	\frac{r^4}{\sqrt{g_{tt}}\sqrt{g_{rr}}}\left(\frac{g_{rr}
-4\pi\eta\Psi^2}{g_{rr}}\right)\frac{\partial}{\partial r}\left[\frac{1}{r^2}\frac{\sqrt{g_{tt}}}{\sqrt{g_{rr}}}\left(\frac{g_{rr}}{g_{rr}
+4\pi\eta\Psi^2}\right)\frac{\partial}{ \partial r}\mathcal{R}\right]-\frac{r^2}{g_{tt}}\frac{\partial^2\mathcal{R}}{\partial t^2}-(\ell+2)(\ell-1)\mathcal{R}=0~.
\end{equation}
where $\ell$ is the angular quantum number of the perturbation. In order to continue, it will prove useful to set
\begin{equation}
	\label{3.12}
	h=\frac{\sqrt{g_{tt}}}{\sqrt{g_{rr}}}\, , \qquad
	A=\left(\frac{g_{rr}}{g_{rr}+4\pi\eta\Psi^2}\right)\, ,\qquad
	B=\left(\frac{g_{rr}-4\pi\eta\Psi^2}{g_{rr}}\right)~,\qquad
\end{equation}
thus, simplifying equation (\Ref{3.11}) to
\begin{equation}
	\label{3.13}
	B r^2 h\frac{\partial}{\partial r}\left[\frac{1}{r^2}Ah\frac{\partial}{\partial r}\mathcal{R}\right]-\frac{\partial^2\mathcal{R}}{\partial t^2}-\frac{(\ell+2)(\ell-1)}{r^2}g_{tt}\mathcal{R}=0~.
\end{equation}
By further defining a scalar field
\begin{equation}
	\label{3.14}
	u=\frac{\mathcal{R}}{r}\sqrt{A}~,
\end{equation}
and using the tortoise coordinate transformation $h=dr/dr_*$, we find that the equation that governs the axial perturbations reads
\begin{equation}
	\label{3.15}
	h\frac{\partial}{\partial r}\left[h\frac{\partial}{\partial r}u\right]
-\frac{1}{AB}\frac{\partial^2u}{\partial t^2}-\left[\frac{(\ell+2)(\ell-1)}{r^2}\frac{g_{tt}}{AB}
+\frac{r}{\sqrt{A}}h\frac{\partial}{\partial r}\left(h\frac{\partial}{\partial r}\left(\frac{\sqrt{A}}{r}\right)\right)\right]u=0~.
\end{equation}
If we consider a time dependence of the form $u(r,t)\sim u(r)\exp(i \omega t)$, then (\Ref{3.15}) yields a non-trivial wave equation of the following form
\begin{equation}
	\label{3.16}
	h\frac{d}{dr}\left[h\frac{d}{dr}u\right]+\frac{\omega^2}{AB}u
-\left[\frac{(\ell+2)(\ell-1)}{r^2}\frac{g_{tt}}{AB}+\frac{r}{\sqrt{A}}h\frac{d}{dr}\left(h\frac{d}{dr}\left(\frac{\sqrt{A}}{r}\right)\right)\right]u=0~.
\end{equation}
As such, the corresponding Regge-Wheeler-like potential reads
\begin{equation}
	\label{3.17}
	V(r)=\frac{(\ell+2)(\ell-1)}{r^2}\frac{g_{tt}}{AB}+\frac{r}{\sqrt{A}}h\frac{d}{dr}\left(h\frac{d}{dr}\left(\frac{\sqrt{A}}{r}\right)\right)~.
\end{equation}
It is important to note that by fixing the metric components to those of the BH solution, Eq. (\Ref{3.17}) asymptotes to the standard Schwarzschild Regge-Wheeler effective potential in the limit where $\ell_\eta\rightarrow +\infty$. Another crucial aspect of Eq. (\Ref{3.16}) is the presence of a multiplication factor $1/AB$ to the gravitational perturbation frequency $\omega$, which signifies the existence of a modified speed of GW propagation \cite{Gong:2017kim}.

\begin{figure}[h]
    \centering
    \includegraphics[scale=0.5]{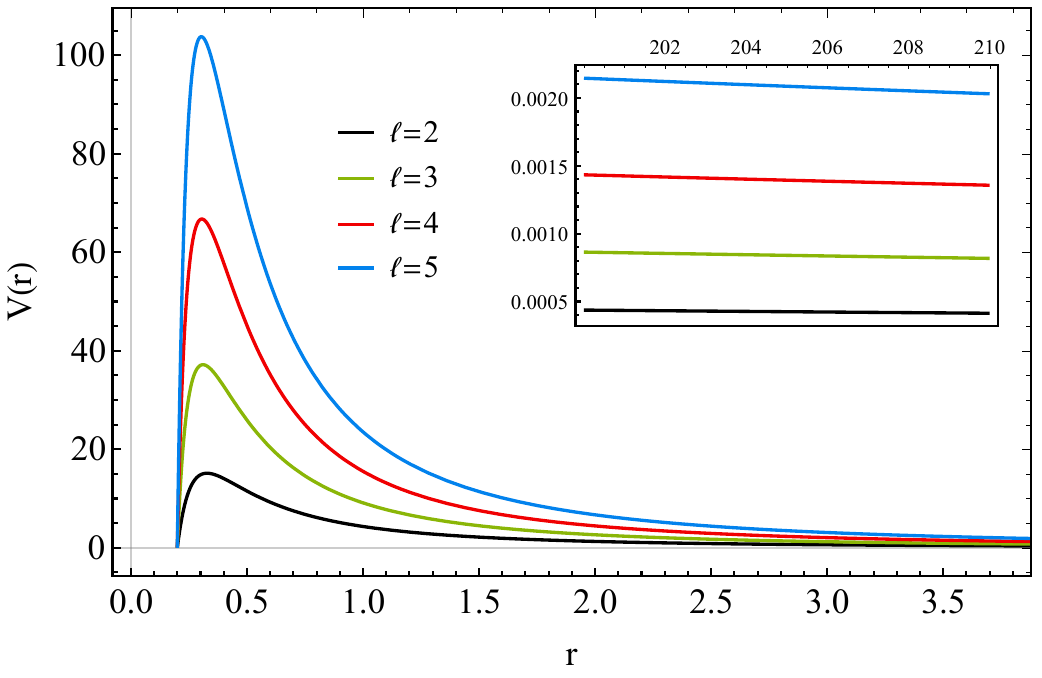}\hspace{0.2 cm}
    \includegraphics[scale=0.485]{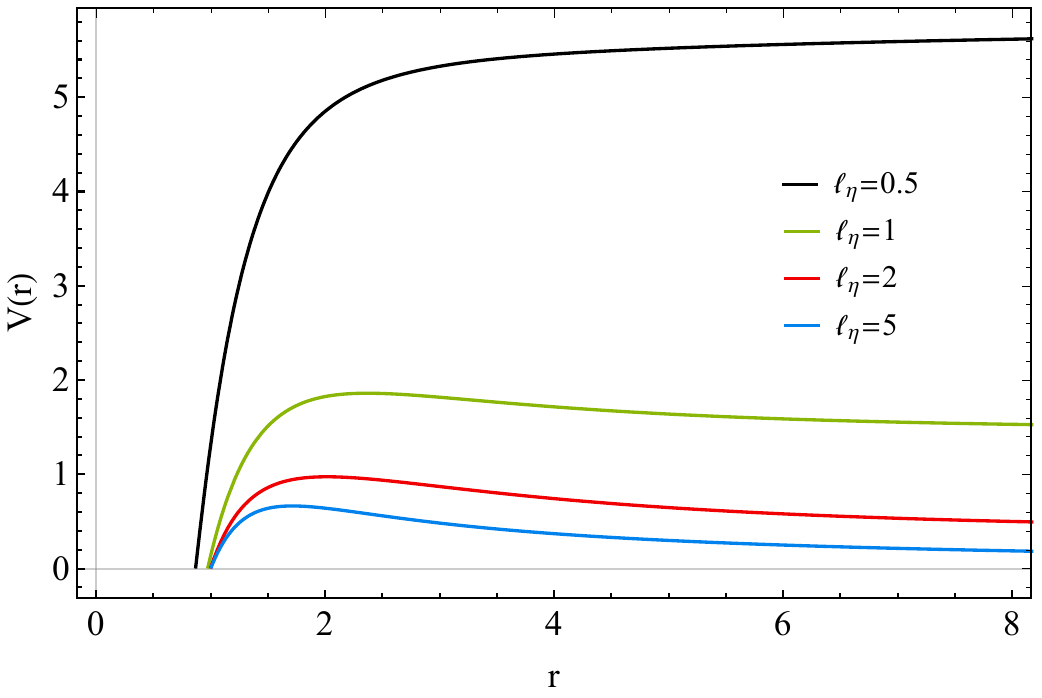}
    \caption{Left: BH effective potential  with $\ell_\eta = 100\,,\,m = 0.1$ and different values of angular numbers $\ell$. Right: BH effective potential with $m = 0.5\,,\,\ell = 2$ and various coupling constants $\ell_\eta$.} \label{fig:V-eff_BH}
\end{figure}

Equation \eqref{3.16} demonstrates that one can reduce the problem of axial gravitational perturbations around the compact object under consideration into a single one-dimensional scattering problem with an effective potential. By applying the procedure outlined above, on the BH solution \eqref{2.3}-\eqref{2.4} we find the corresponding effective potentials a gravitational perturbation induces. An illustration of these potentials for various choices of the angular index of the perturbation $\ell$ is given in Fig. \ref{fig:V-eff_BH}. The effective potential possesses a peak for sufficiently large $\ell_\eta$, at the Schwarzschild limit, which is located arbitrary close to the photon sphere at $r=3m$. For a myriad of BH solutions, Ref. \cite{Cardoso:2008bp} demonstrates that the angular frequency and instability timescale of null geodesics that are trapped in unstable circular orbits at the photon sphere are associated with the oscillation frequency and decay rate of eikonal QNMs. In turn, the existence of such centrifugal potential barrier is responsible for the prompt ringdown and photon sphere QNMs found in the response of a plethora of perturbed BHs \cite{Cardoso:2017soq,Cardoso:2018nvb,Destounis:2018qnb,Liu:2019lon,Destounis:2019omd,Destounis:2020pjk,Destounis:2020yav}. In what follows, we will show that the aforementioned analogy only holds at the GR limit and away from it such duality is broken.

Asymptotically, the potential approaches a constant positive value, a behavior very different from the case of scalar perturbations \cite{Vlachos:2021weq}, but still, one which encodes the effective non-asymptotically-flat nature of spacetime. A similar behavior was also observed in \cite{Bronnikov:2012ch}. More specifically, at spatial infinity, the BH potential at zeroth order yields $V(r\rightarrow\infty) \sim \mathcal{C}+ \mathcal{O}\left(1/r\right)$
where the constant $\mathcal{C}$ depends on the parameters $\ell$ and $\ell_\eta$ and can be calculated only numerically.
A similar asymptotic behavior was also observed in \cite{Abdalla:2018ggo} for the case of vector perturbations in the scalarized BH considered here.

We must note here the dimensionality of various quantities appearing in the following figures, in order to avoid repetition and cluttering on our discussion. According to the geometrized units utilized here, the BH mass $m$ and coupling $\ell_\eta$ have dimensions of length, while the perturbation $u$ is dimensionless, as well as the ratio $m/\ell_\eta$, which makes it an appropriate scale for our analysis. In turn, the effective potential $V(r)$ has dimensionality length to the power of $-2$, as expected, while the frequency $\omega$ has length dimensions.

\section{Time-domain integration scheme}
\label{sec4}

Here, we briefly demonstrate the numerical scheme of time-domain integration, first proposed in \cite{Gundlach:1993tp}, which yields the temporal response of a metric perturbation as it propagates on a fixed background spacetime. By defining $u(r_\ast,t)=u(i\Delta r_\ast,j\Delta t)=u_{i,j}$, $V(r(r_*))=V(r_\ast)=V(i\Delta r_\ast)=V_{i}$, $A(r_\ast)=A(i\Delta r_\ast)=A_i$ and $B(r_\ast)=B(i\Delta r_\ast)=B_i$ equation \eqref{3.16} takes the form
\begin{align}\label{time_ev}
    \frac{u_{i+1,j} - 2u_{i,j} + u_{i-1,j} }{\Delta r^2_\ast} - \frac{1}{A_i B_i} \frac{ u_{i,j+1} - 2u_{i,j} + u_{i,j-1} }{\Delta t^2} - V_i \,u_{i,j} = 0\;.
\end{align}
Then, by using as initial condition a Gaussian wave-packet of the form $\psi(r_\ast,t) = \exp\left[ -(r_\ast-c)^2/(2\sigma^2) \right]$ and $\psi(\rho_\ast,t<0) = 0$, where $c$ and $\sigma$ correspond to the median and width of the wave-packet, we can derive the time evolution of $u$
\begin{align}
    u_{i,j+1} = A_i B_i \,\Delta t^2 \left( \frac{u_{i+1,j} - 2u_{i,j} + u_{i-1,j} }{\Delta r^2_\ast} - V_i u_{i,j} \right) + 2 u_{i,j} - u_{i,j-1}~,\label{recursive-relation}
\end{align}
where the Courant-Friedrichs-Lewy (CFL) condition requires $\Delta t/\Delta r_\ast < 1/(A_i B_i)$. To calculate the precise values of the potential $V_i$, we integrate numerically the equation for the tortoise coordinate and then solve with respect to the corresponding radial coordinate. Moreover we require the vanishing of perturbations at radial infinity by imposing reflective boundary conditions $u_{i_{max}, j} = 0$, since our solution is effectively asymptotically AdS \cite{Cardoso:2001bb,Berti:2009kk}.
For further details regarding the numerical scheme and its convergence see Appendix \ref{appB}.

\section{Evolution of axial gravitational perturbations}
\label{sec5}

\begin{figure}[H]
	\centering
	\includegraphics[scale=0.6]{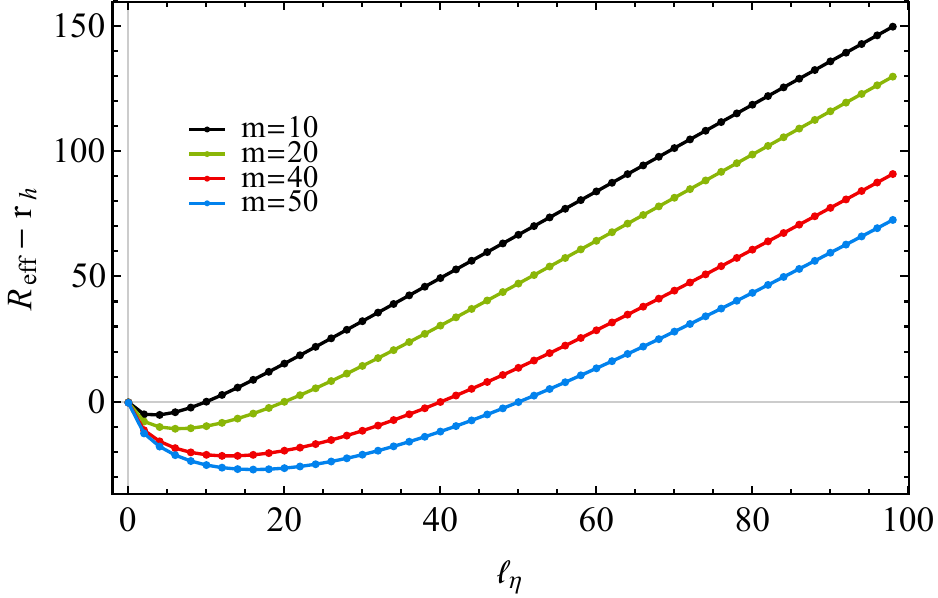}
	\caption{The difference between the effective cosmological and event horizon radius as a function of $m$ and $\ell_\eta$, where $R_{\rm eff} = \sqrt{3}\,\ell_{\eta}$. For $m<<\ell_\eta$ we have $r_h << R_{\rm eff}$, for $m\simeq\ell_\eta$ we have $r_h\simeq R_{\rm eff}$, while for $m>>\ell_\eta$ we have $r_h >> R_{\rm eff}$. The minima of the curves occur at approximately $m/\ell_\eta\simeq3$ and indicate the points in the parametric space $(m,\ell_\eta)$ for which $r_h >> R_{\rm eff}$.} \label{AdSvshorizon}
\end{figure}

Regardless of the fact that the spacetime utilized here does not describe a BH immersed in a Universe with a negative cosmological constant, it is meaningful to compare it with Schwarzschild-AdS BHs, since $\ell_\eta$ introduces an effective cosmological scale to the geometry considered. It what follows, we will adopt the categorization from \cite{Horowitz:1999jd} regarding AdS BHs determined by two dimensionful parameters, namely the AdS radius $r=R_{AdS}$ and the event horizon radius $r=r_h$. The BH solution in the present study depends also on two parameters: $m$ and $\ell_{\eta}$. The value of mass controls the position of the event horizon $r_{h}$ (and consequently of the photon sphere) and $\ell_{\eta}$ dictates the value of the effective cosmological radius $r=R_{\rm eff}$. However, one key difference between the two solutions is that the first is a `bald' BH embedded in an AdS Universe, whereas the scalarized solution is `dressed' with scalar hair whose existence creates the effective AdS-like asymptotics. In this sense, the parameter $\ell_\eta$ controls the strength of the scalar hair and as a consequence the value of $R_{\rm eff}$. In our case, the effective cosmological radius is given by $R_{\rm eff} = \sqrt{3}\,\ell_{\eta}$. One may categorize BHs in an AdS Universe \cite{Horowitz:1999jd} as (i) small size BHs with $r_h << R_{AdS}$, (ii) intermediate size BHs with $r_h\sim R_{AdS}$ and (iii) large BHs with $r_h >> R_{AdS}$. We utilize a similar classification to distinguish between small ($r_h << R_{\rm eff}$), intermediate ($r_h\sim R_{\rm eff}$) and large hairy BHs ($r_h >> R_{\rm eff}$) (see Fig. \ref{AdSvshorizon}).

\begin{figure}[H]
	\centering
	\includegraphics[scale=0.48]{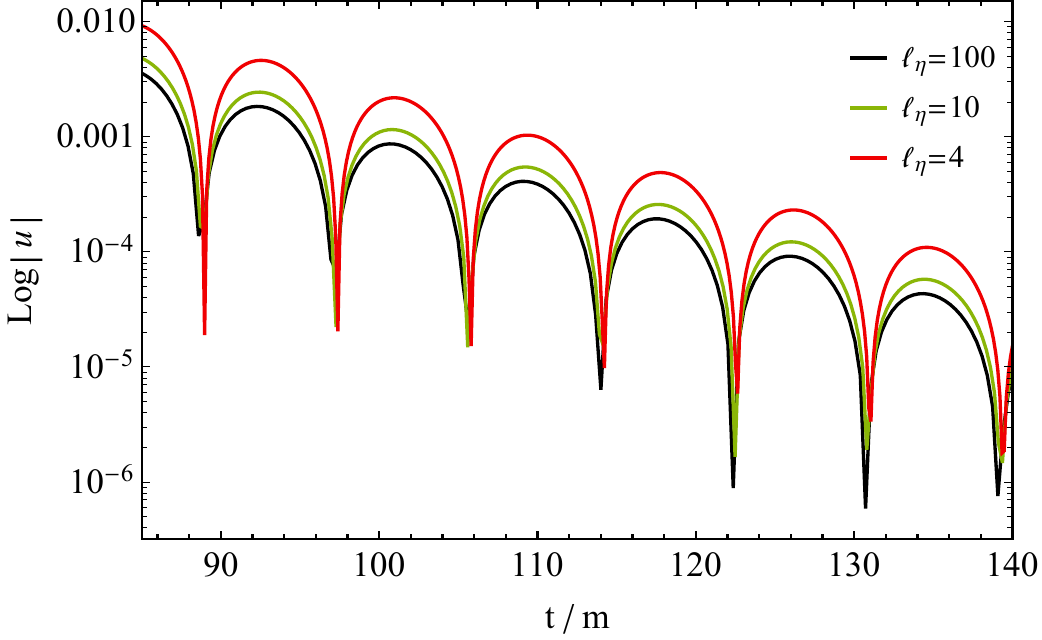}\hskip 1ex
	\includegraphics[scale=0.44]{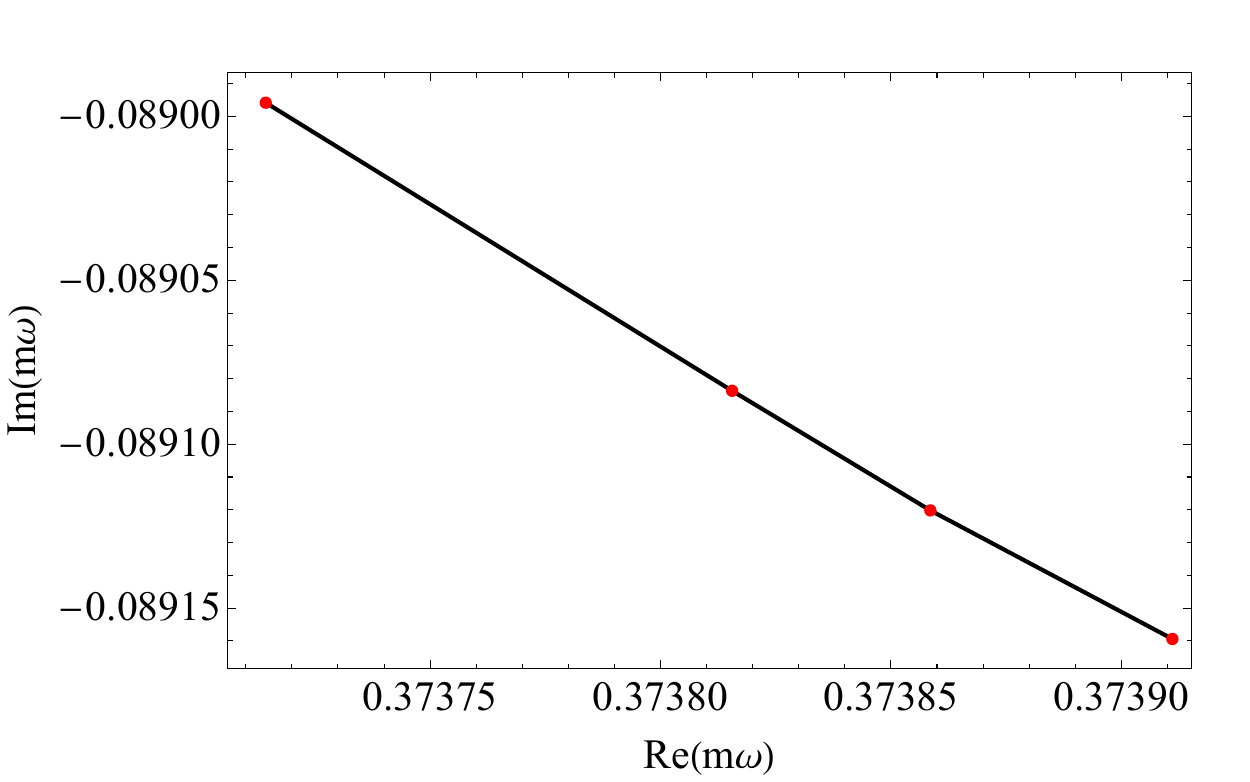}
	\caption{Left: Oscillatory prompt response of axial gravitational perturbations with $\ell=2$ of the BH considered with $m=0.1$ and varying $\ell_\eta$. Right: Fundamental $\ell=2$ modes extracted from the prompt ringing phase of the BH considered with $m=0.1$ and $\ell_\eta=4,\,10,\,50,\,100$ from left to right.} \label{QNM_convergence}
\end{figure}

In what follows, we apply the numerical procedure outlined above, to calculate the temporal response of axial gravitational perturbations on the BHs of the above categories. In the following figures we obtain the perturbation response at a position arbitrarily close to the event horizon $r-r_h=10^{-5}$, though we have checked that the same results are obtained if we calculate the response at any position outside the event horizon. Furthermore, we have performed some typical tests to ensure the validity of the integration method. Specifically, we have calculated the response of gravitational perturbations on the BH considered here, in the limit where $\ell_\eta\rightarrow\infty$, where the effect of the scalar hair is suppressed, (we have chosen $\ell_\eta=1000$ though even for $\ell_\eta=10$ the potential $V(r)$ converges to the Regge-Wheeler one). By using the Prony method \cite{Berti:2007dg} we can extract the spectral content from the temporal response at the large coupling limit and investigate if the modes extracted solely from the prompt ringdown agree with the standard axial gravitational QNMs of Schwarzschild BHs \cite{Chandrasekhar}. In Fig. \ref{QNM_convergence} we show the prompt ringing of small BHs for $\ell=2$ gravitational perturbations. We only consider the case where the BH mass is $m=0.1$ in order to obtain a clear ringing phase, since for the range of couplings we considered from $4$ to $100$ the echo timescales are very large and the extraction of QNMs from the prompt ringdown is possible. Figure \ref{QNM_convergence} indicates that decreasing $\ell_\eta$ has a menial effect on the spectrum while for $\ell_\eta=100$ the extracted mode asymptotes to the fundamental Schwarzschild QNM with accuracy $\sim0.1\%$.

For completeness, we have further calculated the instability timescale of null geodesics (Lyapunov exponents) at the photon sphere \cite{Cardoso:2008bp} and found that at the GR limit the correspondence between null geodesics and eikonal (large $\ell$) QNMs still holds. This is expected since at this limit the BH approaches Schwarzschild and GWs propagate with the speed of light. Therefore this analysis serves as another validity test of our numerical results and justifies the existence of a modified propagation speed of GWs. In fact, for the case with $m=0.1$, $\ell_\eta=100$ ($m/\ell_\eta=10^{-3}$) the instability timescale of null geodesics and the extracted decay rate of the fundamental QNM for $\ell=10$ (approximately eikonal) axial perturbations have a percentage difference of less than $0.5\%$. On the other hand, when $\ell_\eta$ is not large enough then the propagation speed of GWs is modified, in accordance with Eq. \eqref{3.16}, and this leads to a significant inconsistency between null geodesics and eikonal fundamental QNMs, as expected. For example, by choosing $m=5$, $\ell_\eta=1$ ($m/\ell_\eta=5$) the instability timescale of null geodesics and the extracted decay rate of the fundamental QNM for $\ell=10$ (approximately eikonal) axial perturbations have a percentage difference $\sim50\%$. Therefore we conclude that the fundamental eikonal QNMs are not always associated with null geodesics at the spacetime, as it was also shown in \cite{Konoplya:2017wot}.

\begin{figure}[H]
	\centering
	\includegraphics[scale=0.485]{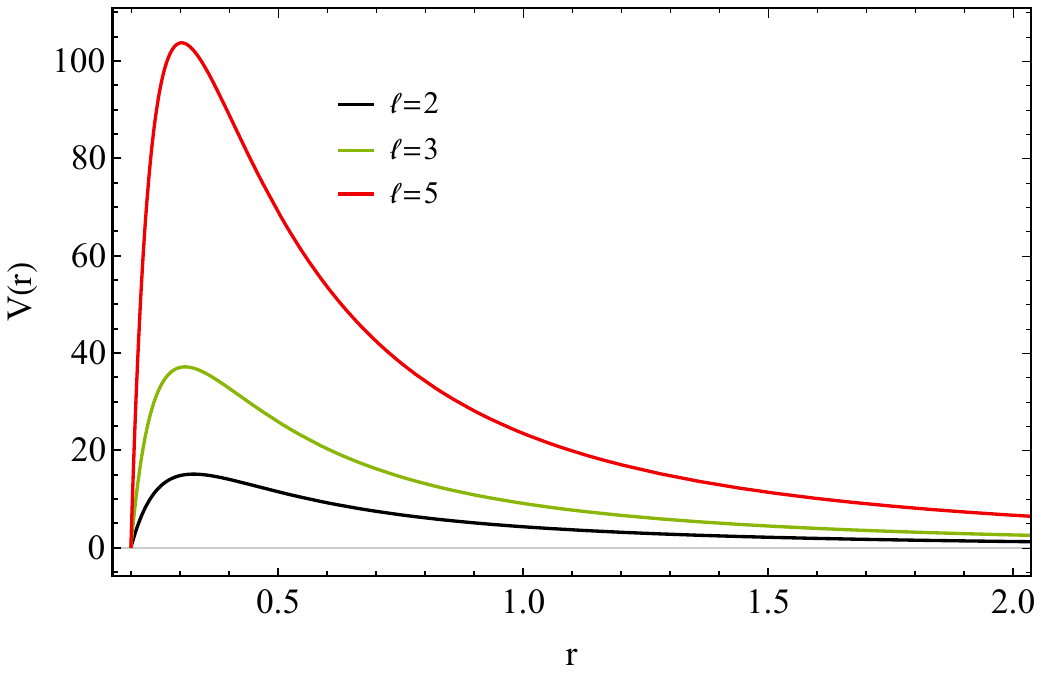}\hspace{0.2 cm}
	\includegraphics[scale=0.5]{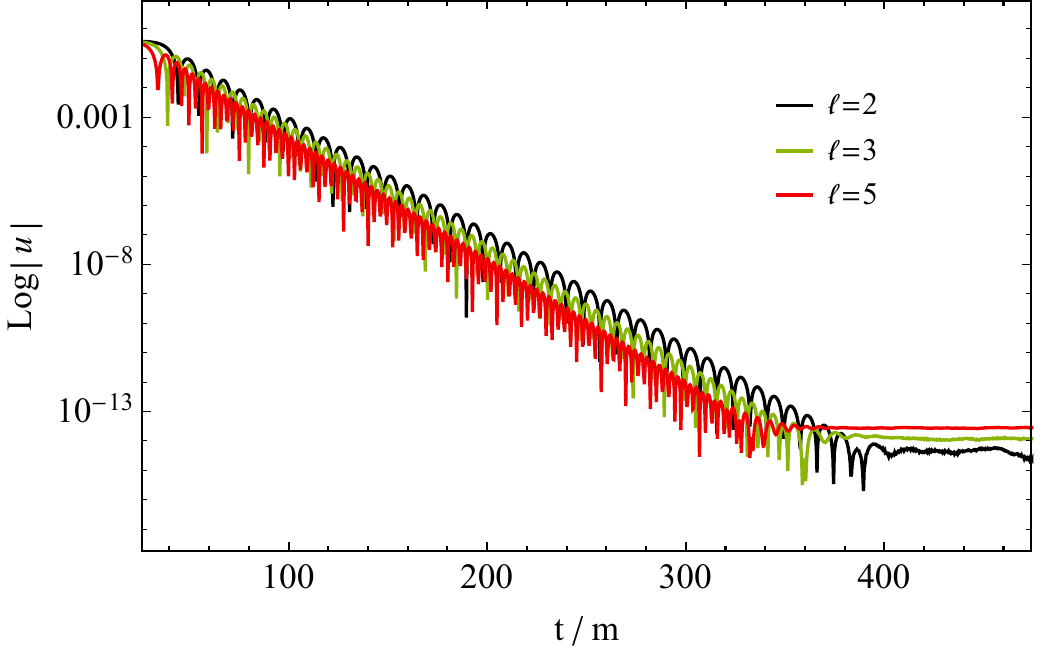}
	\caption{Left: Effective potential of axial gravitational perturbations, with varying $\ell$, of the scalarized BH with $\ell_\eta = 100$, $m = 0.1$. The parametric region considered here corresponds to small BHs in the sense that $r_h<<R_{\rm eff}$ ($m/\ell_\eta\sim\mathcal{O}(10^{-3})$). Right: Response of axial gravitational perturbations with respect to the effective potential of BHs considered at the left subfigure.} \label{fig:2}
\end{figure}

Figure \ref{fig:2} presents the temporal response of the BH against axial perturbations in the small ratio regime $m<<\ell_\eta$ ($m/\ell_\eta\sim\mathcal{O}(10^{-3})$). The initial quasinormal ringdown is quite similar to that of a Schwarzschild BH. Such behavior is expected and can be attributed to the high value of $\ell_\eta$ relative to the mass. Perturbations with higher angular number $\ell$ decay faster and with higher frequency since more energy is carried away from the photon sphere. This phenomenon is expected since similar behavior appears for gravitational perturbations and QNMs in Schwarzschild BHs \cite{Berti:2009kk}. The late time behavior however, shows that, instead of a power-law cutoff, the field settles to a constant value which is related to the asymptotic value that the effective potential acquires (see Fig. \ref{fig:V-eff_BH}) and the expectancy of late-time echoes. The eventual late-time tail should be more evident for large BHs since echoes will be washed out rapidly at the event horizon.

\begin{figure}[H]
    \centering
    \includegraphics[scale=0.48]{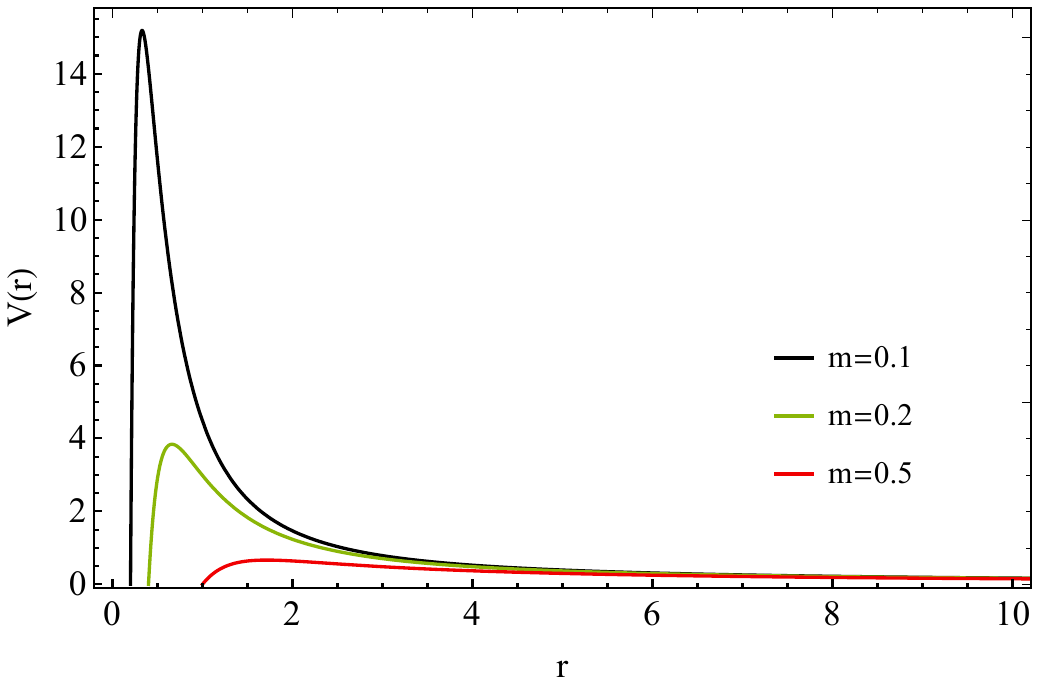}\hspace{0.2 cm}
    \includegraphics[scale=0.51]{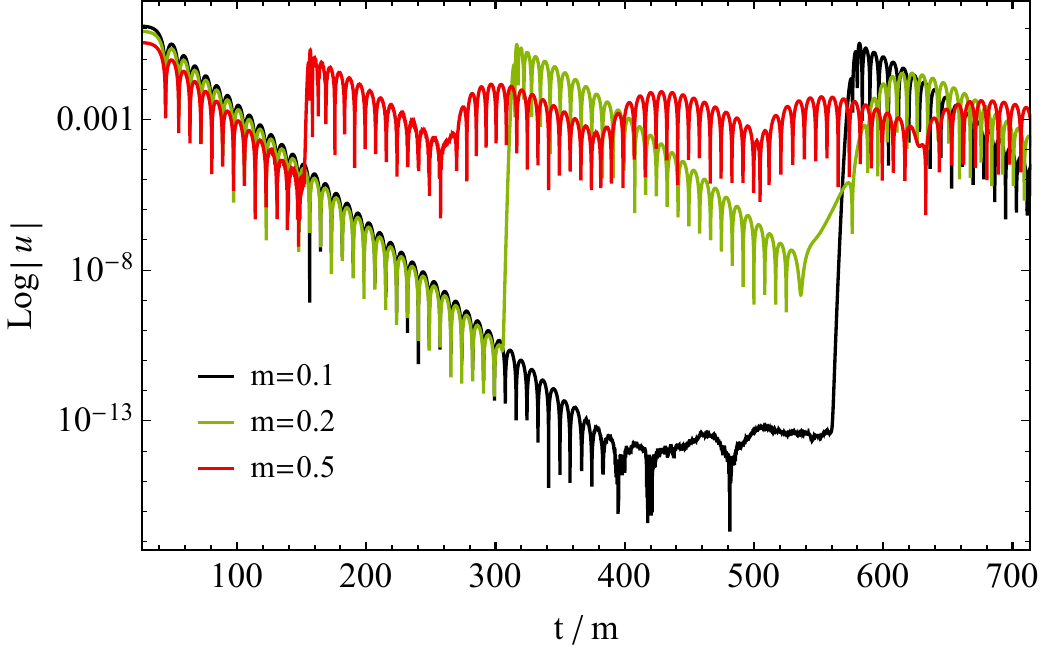}
    \caption{Left: Effective potential of an $\ell=2$ gravitational perturbation of the scalarized BH with $\ell_\eta=5$ for varying mass $m$. The cases we consider here correspond to intermediate size BHs with $r_h\sim R_{\rm eff}$ ($m/\ell_\eta\sim\mathcal{O}(10^{-1})$). Right: Response of axial gravitational perturbations with respect to the effective potential of BHs considered at the left subfigure.} \label{fig:3}
\end{figure}

In Fig. \ref{fig:3} the evolution of perturbations for $m<\ell_\eta$ ($m/\ell_\eta\sim\mathcal{O}(10^{-1})$) is displayed. The most obvious effect one observes is the emergence of echoes following the initial quasinormal ringdown. In this parametric region the relation between the mass of the BH and the coupling $\ell_\eta$ becomes more transparent. By keeping $\ell_\eta$ fixed and increasing the mass, perturbations will have to travel a shorter distance between the photon sphere and the effective AdS boundary induced by the scalar field leading to repetitions in the signal which appear in shorter timescales. Analogously, similar behavior is obtained when one keeps the mass fixed and decreases the coupling. This pattern was also observed in \cite{Vlachos:2021weq} for the case of scalar perturbations though test scalar fields travel with the speed of light, in contrast to axial gravitational waves in our analysis which have a variable propagation speed (see Eq. \eqref{3.16}). This means that a null geodesic analysis, similar to that in \cite{Cardoso:2016rao,Cardoso:2016oxy,Cardoso:2019rvt} where the echo timescales are approximated by the time that light takes to travel from a boundary to the photon sphere and back, is rendered pointless. Our case is much more intricate since one cannot consider null geodesics anymore but rather has to analyze waves traveling in a dispersive medium with varying propagation speed in different regimes. We have performed a trivial null geodesic analysis and the results we obtained are expected, that is for large $\ell_\eta$ the propagation speed of GW approaches the one of light and the echo timescales can be properly approximated, while as the coupling decreases the echo timescales predicted by null geodesics are completely inconsistent with the actual timescales of echoes obtained by our numerical integration. Nevertheless, such investigation reinforces our discussion regarding the existence of a modified GW speed of propagation.

\begin{figure}[H]
    \centering
    \includegraphics[scale=0.48]{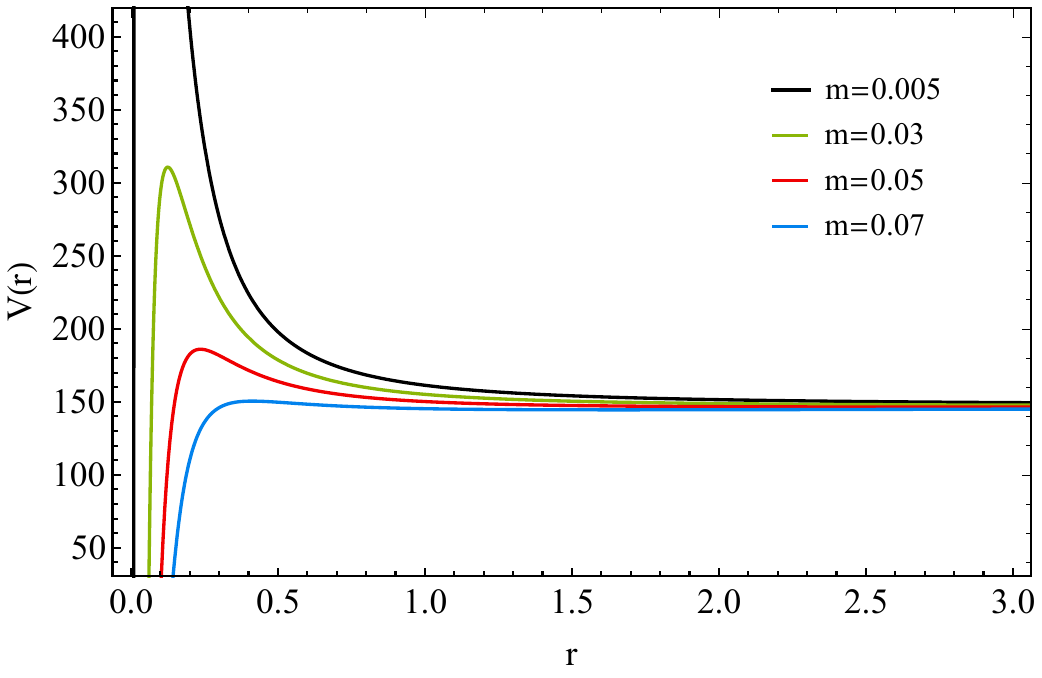}\hspace{0.2 cm}
    \includegraphics[scale=0.495]{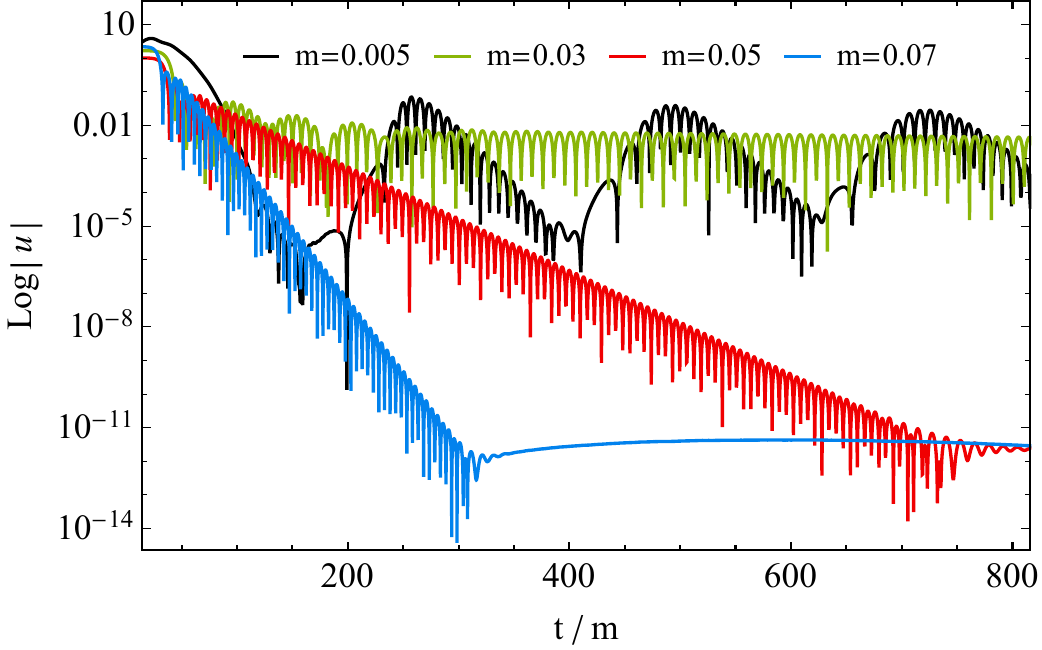}
    \caption{Left: Effective potential of an $\ell=2$ gravitational perturbation of the scalarized BH with $\ell_\eta=0.1$ for varying mass $m$. Right: Response of axial gravitational perturbations with respect to the effective potential of BHs considered at the left subfigure.} \label{fig:4}
\end{figure}

To obtain a complete picture regarding the effect of the ratio $m/\ell_\eta$ on the BH's response to fluctuations we have plotted in Fig. \ref{fig:4} the time evolution of perturbations for a wide range of masses keeping $\ell_\eta$ fixed. As $m$ grows the echoes are replaced by quasinormal oscillations, while further increment of the mass leads to a single quasinormal ringdown followed by a late-time tail. We conclude that this behavior stems from the shape of the effective potential which decreases in amplitude as $m$ increases. This leads to an increasingly smaller region where trapped modes, which lead to echoes, can occur, and thus the quasinormal ringing of the BH dominates over the echoes which are quickly suppressed.

\begin{figure}[H]
    \centering
    \includegraphics[scale=0.47]{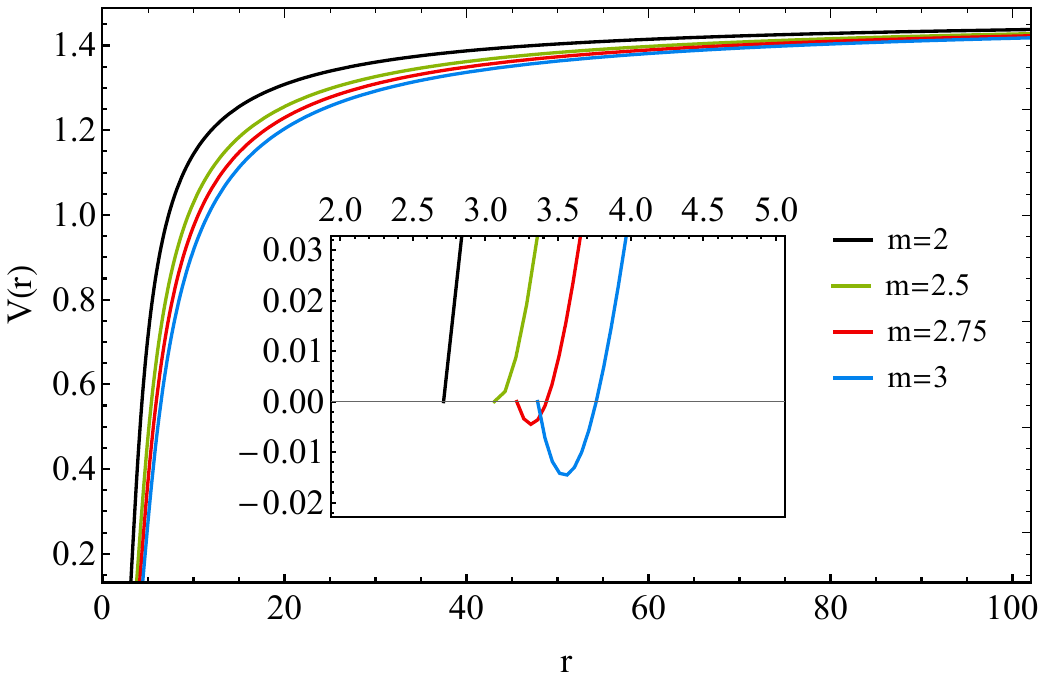}\hspace{0.2 cm}
    \includegraphics[scale=0.49]{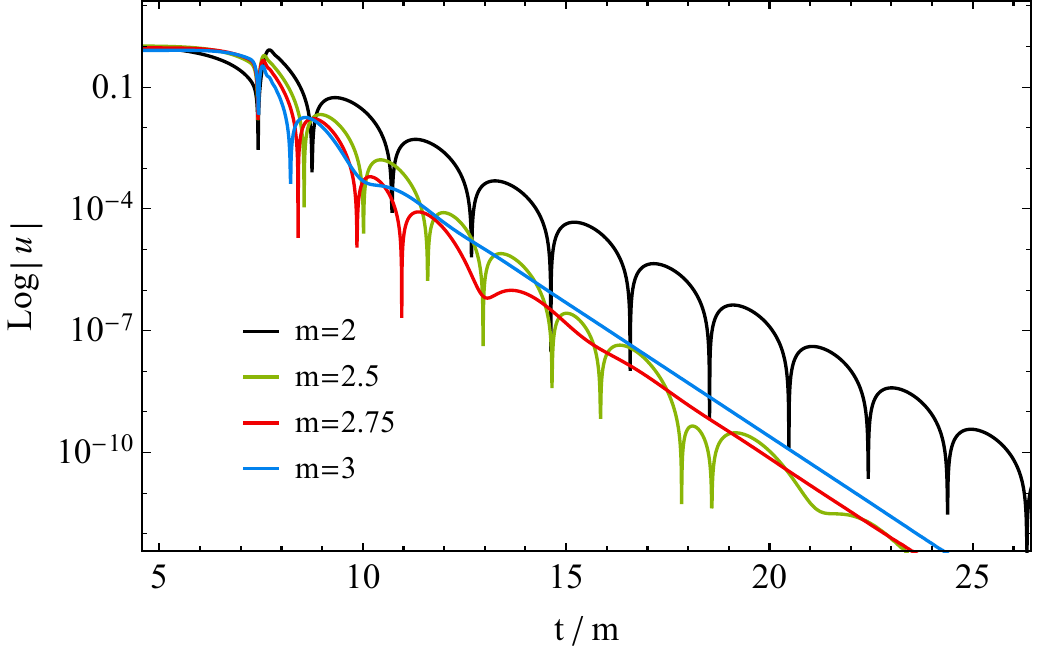}
    \caption{Left: Effective potential of an $\ell=2$ gravitational perturbation of the scalarized BH with $\ell_\eta=1$ for varying mass $m$. The cases we consider here correspond to intermediate size BHs with $r_h\sim R_{\rm eff}$ ($m/\ell_\eta\sim\mathcal{O}(10^{0})$). Right: Response of axial gravitational perturbations with respect to the effective potential of BHs considered at the left subfigure. The responses were shifted in time for illustration purposes.}
    \label{fig:5}
\end{figure}

\begin{figure}[H]
    \centering
    \includegraphics[scale=0.485]{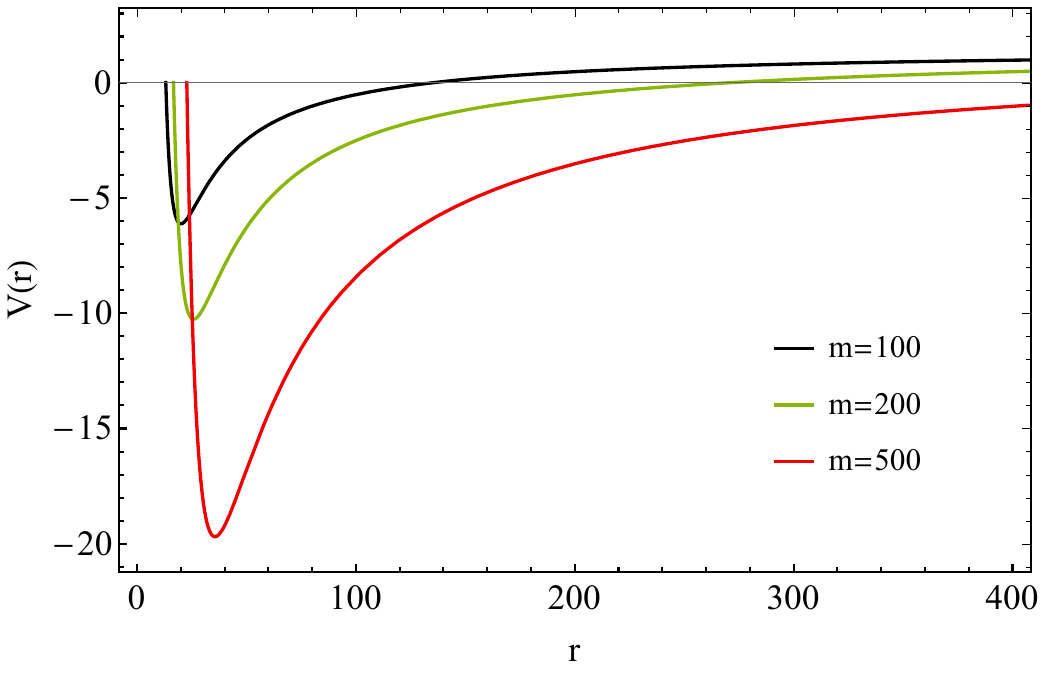}\hspace{0.2 cm}
    \includegraphics[scale=0.49]{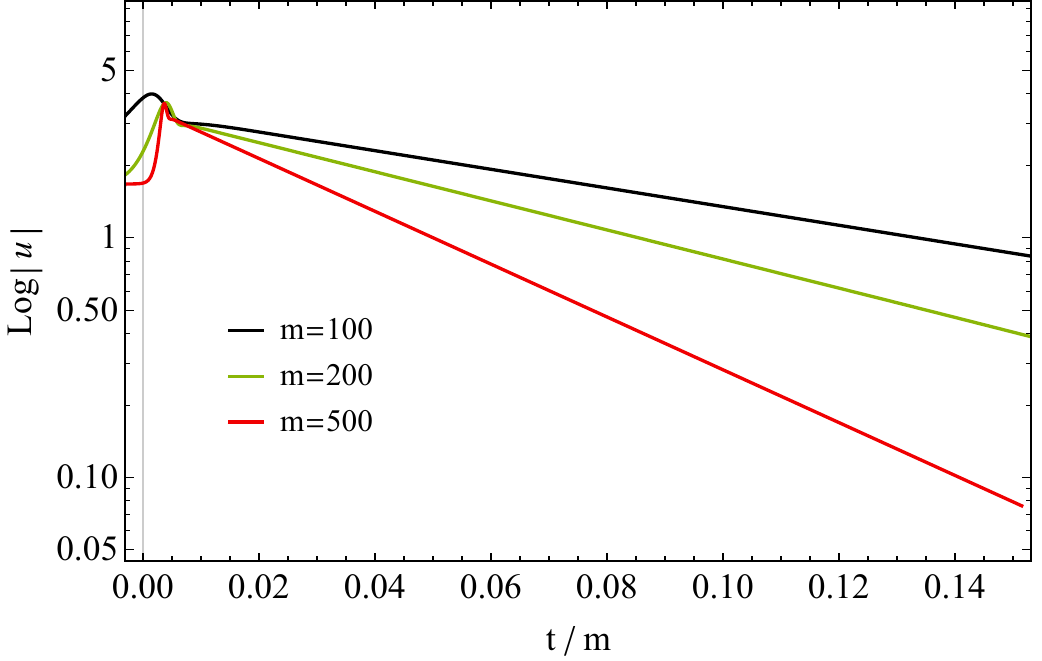}
    \caption{Left: Effective potential of an $\ell=1$ gravitational perturbation of the scalarized BH with $\ell_\eta=5$ for varying mass $m$. The cases we consider here correspond to large BHs with $r_h >> R_{\rm eff}$ ($m/\ell_\eta\sim\mathcal{O}(10^{2})$). Right: Response of axial gravitational perturbations with respect to the effective potential of BHs considered at the left subfigure. A shift in time was applied on the responses for illustration purposes.}
    \label{fig:6}
\end{figure}

When the mass becomes proportional ($m/\ell_\eta\sim\mathcal{O}(10^0)$) or significantly larger than $\ell_\eta$ ($m/\ell_\eta\sim\mathcal{O}(10^2)$), negative wells develop in the effective potential in the vicinity of the event horizon (see Figs. \ref{fig:5}, \ref{fig:6}). Despite the negative well formation, the time-domain profiles show an exponential decay of the signal without any indication of a linear instability. On the contrary, more massive objects lead to signals with shorter quasinormal ringing stages, due to the absence of a photon sphere peak, and with faster decay rates even though the corresponding effective potentials develop even deeper negative wells. The exponential nature of the eventual late-time behavior of perturbations is related to the effective AdS asymptotics of our spacetime which requires the imposition of reflective boundary conditions at infinity and is in agreement with what occurs in perturbations of AdS BHs \cite{Holzegel:2011uu}. The tail in these cases appears because echoes are subdominant and vanish very rapidly at the event horizon, thus the asymptotic behavior is probed faster. We expect that even perturbations of the small BHs in study will eventually possess an exponential tail but at much later times which our numerical scheme cannot probe.

\begin{figure}[H]
    \centering
    \includegraphics[scale=0.46]{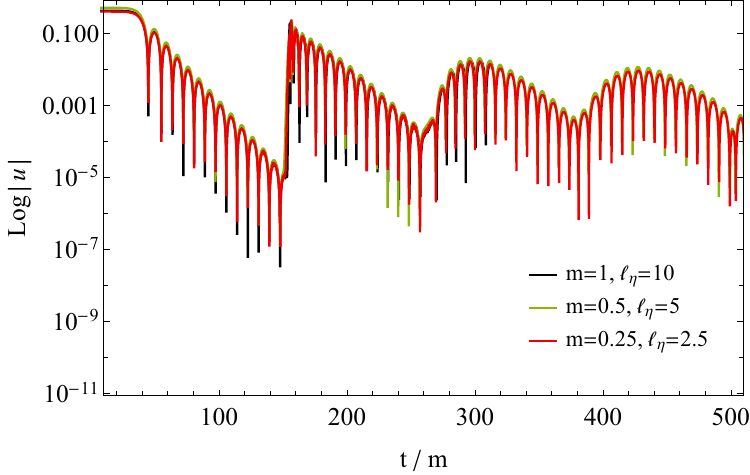}\hspace{0.1cm}
    \includegraphics[scale=0.33]{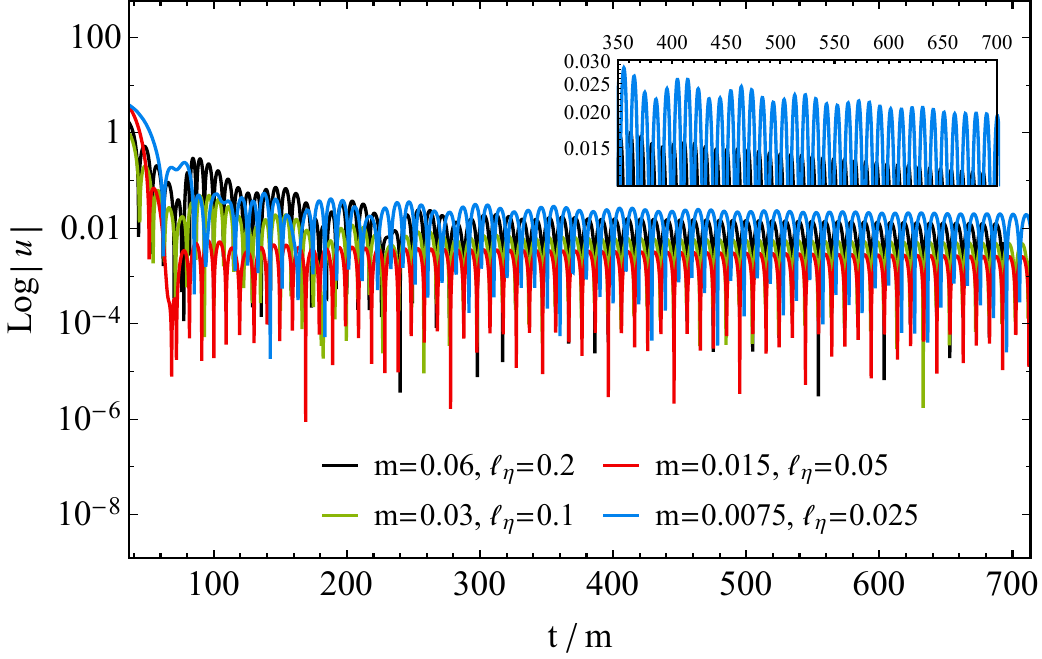}\hspace{0.1cm}
    \includegraphics[scale=0.46]{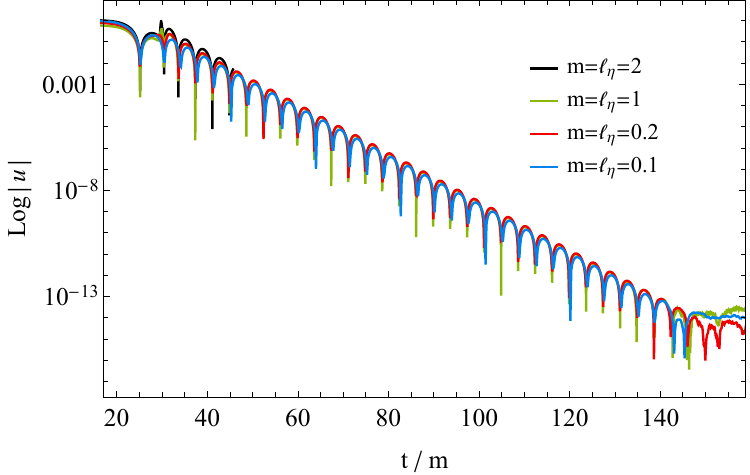}
    \caption{Time evolution of an $\ell=2$ gravitational perturbation of the scalarized BH with fixed ratio $m/\ell_\eta\simeq 0.1$ (left), $m/\ell_\eta\simeq 0.3$ (middle) and $m/\ell_\eta=1$ (right).}\label{movl}
\end{figure}

From the above, we conclude that the scalarized BH spacetime is modally stable under axial gravitational perturbations, where the qualitative features of the response depend solely on the ratio $m/\ell_\eta$. In Fig. \ref{movl} we demonstrate the above statement for the case of intermediate size BHs. Our numerics show that a similar analogy occurs irregardless of the BH's size. Even though the source of echoes in our BH is related to the asymptotics of spacetime, and not to the nature of the near-horizon structure, our results are in accordance with perturbations in wormholes with decreasing throat radii \cite{Liu:2020qia} and black-bounce models, which transition from regular BHs to wormholes \cite{Churilova:2019cyt}.

\section{Conclusions}
\label{sec6}

In this work we studied static and spherically symmetric solutions of a Horndeski subclass which includes a massless scalar field non-minimally coupled to the Einstein tensor. Such theory admits an exact BH solution `dressed' with scalar hair whose existence induces an effective negative cosmological constant even though the BH does not reside in an AdS Universe. We have studied the modal stability of such solutions under axial gravitational perturbations, with time evolution techniques, and complementary QNM extraction, that solve the linearized gravitational wave equation. Our results designate that the BH under study is linearly stable against axial perturbations, with decaying temporal responses akin to ringdown waveforms. The qualitative features of the ringdown waveform depend solely on ratio of the two available parameters of spacetime, namely the BH mass $m$ and non-minimal coupling strength $\ell_\eta$. We have further demonstrated that as $m/\ell_\eta$ increases, we have   gravitational-wave ringdown transitions between three distinct response patterns,  namely a state with a typical quasinormal ringdown ($m/\ell_\eta\lesssim 10^{-2}$), an intermediate long-lived state which exhibits gravitational-wave echoes ($10^{-2}\lesssim m/\ell_\eta\lesssim 10^{-1}$) and a state where the ringdown and echoes are depleted rapidly to give turn to an exponential tail ($m/\ell_\eta\gtrsim 10^{-1}$).

Regardless that our findings point towards linear stability, we only considered the axial sector of gravitational fluctuations. In generality, one must investigate the polar sector of gravitational perturbations as well in order for a complete stability analysis to be established. This extension can be extremely challenging with what regards the achievement of writing the perturbation equation into a one-dimensional Zerilli-like equation and the stability of spacetime itself, since the polar degrees of freedom generically couple to the scalar hair in scalar-tensor theories. A first step towards the aforementioned direction is the consideration of radial perturbations which are a good proxy to polar ones \cite{Torii:1996yi,Zou:2020zxq,Blazquez-Salcedo:2018jnn,Blazquez-Salcedo:2022omw}. Radial perturbations can also couple the scalar field with the metric components, thus can serve as more sensible probe to the overall linear stability of the hairy BHs under consideration.
	
Besides dealing with temporal evolution techniques, another interesting direction would be a complete frequency domain analysis of axial and polar gravitational QNMs which is still lacking in the particular family of BH solutions, in a similar manner as in Refs. \cite{Minamitsuji:2014hha,Dong:2017toi} where scalar QNMs have been discussed. Furthermore, since the BH geometry in study possesses a propagation speed for GWs that differs from that of light, it will be paramount to investigate potential observational imprints in order to disentangle possible degeneracies between GW phase modifications and environmental effects and avoid misinterpreting GWs in modified gravity with strongly-lensed GR GWs \cite{Cardoso:2021wlq,Ezquiaga:2022nak}.

Finally, in a recent analysis \cite{Deffayet:2021nnt}, a class of mechanical models were studied, where a canonical degree of freedom interacts with another one with a negative kinetic term, i.e. with a ghost. Surprisingly, it was shown that the classical motion of the system is completely stable for all initial conditions, even though one would expected that such system to be unstable due to the presence of a ghost field. In our case, we have dealt with a conceptually analogue system, consisting of a scalarized BH for which
the kinetic energy of the scalar hair can be positive or negative (first degree of freedom) provided that the strength of the non-minimal coupling to the Einstein tensor has the opposite sign (second degree of freedom), being attractive of repulsive respectively. Regardless of the case, we find that the BH is stable under axial perturbations, thus providing an illustration that the classical mechanics analysis in \cite{Deffayet:2021nnt} can potentially apply to BH physics.

\appendix

\section{Solution of the angular differential equation}\label{appA}
In this appendix, we present the solution of the angular part of the differential equation (\Ref{3.10}). The corresponding differential equation is:
	\begin{equation}
		\label{a1}
		\sin^3\theta\frac{d}{d\theta}\left[\frac{1}{\sin^3\theta}\frac{d\mathcal{S}(\theta)}{d\theta}\right]+\mathcal{A}\, \mathcal{S}(\theta)=0~,
	\end{equation}
	where $\mathcal{A}$ is the separation constant.
	By performing a change of variables of the form $x=\cos\theta$ we obtain the following differential equation:
	\begin{equation}
		\label{a2}
		(1-x^2)\mathcal{S}''+2x\mathcal{S}'+\mathcal{A}\mathcal{S}=0~.
	\end{equation}
	Note that this is very similar to the Legendre differential equation albeit with one sign change. This differential equation is called the ultraspherical or Gegenbauer differential equation. There exist three alternate forms of the equation that yield the same result. We are going to show the two we are interested in here.
	\\
	
	{\noindent\bf\textit{First form:}}

		 \begin{equation}
			\label{a3}
		(1-x^2)\mathcal{S}''-2(m+1)x\mathcal{S}'+(\ell-m)(\ell-m+1)\mathcal{S}=0~.
		\end{equation}
		The first form has the following solutions
		\begin{equation}
			\label{a4}
			\mathcal{S}=(x^2-1)^{-m/2}\left[C_1P^m_{\ell}(x)+C_2\mathcal{Q}^m_{\ell}(x)\right]~,
		\end{equation}
		where $P^m_\ell(x)$ and $\mathcal{Q}^m_\ell(x)$ are the Legendre functions of the first and second kind, respectively. Note that $m=-2$.
		\\
		
	{\noindent\bf\textit{Second form:}}
	
	\begin{equation}\label{a5}
		(1-x^2)\mathcal{S}''-(2n+1)x\mathcal{S}'+k(k+2n)\mathcal{S}=0~.
	\end{equation}
The second form has the following solutions
		\begin{equation}
			\label{a6}
			\mathcal{S}=(x^2-1)^{(1-2n)/4}\left[C_1P^{n-1/2}_{-1/2+k+n}(x)+C_2\mathcal{Q}^{n-1/2}_{-1/2+k+n}(x)\right]~.
		\end{equation}
		Note that $n=-3/2$. If $-1/2+k+n$ is an integer, then this solution yields the Gegenbauer polynomials, $C^n_k$. Equating (\Ref{a3}) with (\Ref{a4}), yields $-1/2+k+n=l\rightarrow k=\ell+2$, and thus we are left with the solution of
	\begin{equation}
		\label{a7}
	 \mathcal{S}=C^{-3/2}_{\ell+2}(\theta)~.
	\end{equation}
	Therefore, the differential equation (\Ref{a2}) yields the separation constant $\mathcal{A}=(\ell+2)(\ell-1)$~.
	
\section{Convergence tests}\label{appB}
\begin{figure}[h]
	\includegraphics[scale=0.4]{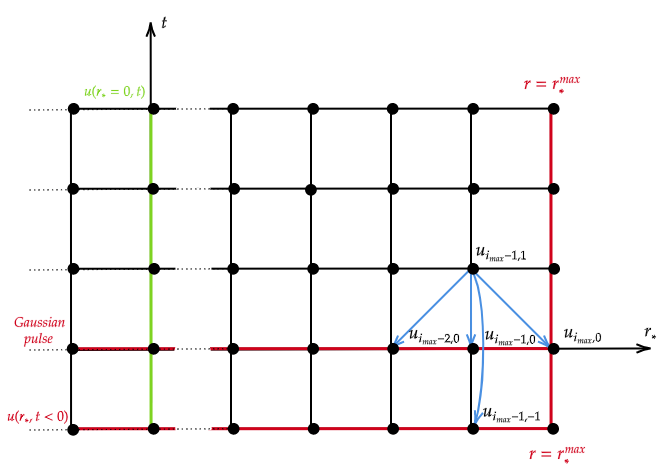}
	\caption{Diagram of the numerical grid in the $(r_* - t)$ plane. The points on the red lines are determined through the boundary and initial conditions. The points on the green line $u(r_*=0,t)$ are the results shown in this paper and they correspond to $r=1.00001\,r_{h}$. The rest of the grid points are calculated through the recursive relation \eqref{recursive-relation} starting from the point $u_{i_{max}-1,1}$. The blue arrows demonstrate a graphical depiction of the time evolution equation \eqref{recursive-relation} for the point $u_{i_{max}-1,1}$.}
	\label{fig:numgrids}
\end{figure}

Here, we discuss in depth our numerical scheme which is briefly analyzed in Section \ref{sec4}. The essential equations in play are Eq. \eqref{time_ev} and \eqref{recursive-relation}, together with the CFL condition and the vanishing of perturbations at radial infinity. In terms of the tortoise coordinate $r_*$, we observe that when $r$ tends to infinity, $r_*$ tends to a finite constant which we denote as $r_*^{max}$. The implications of the behavior of $r_*$ are twofold: firstly, the reflective boundary condition in terms of $r_*$ takes the form $u(r_\ast^{max},t) = u_{i_{max}, j} = 0$ and secondly, our region of interest in the $(r_* - t)$ diagram lies on the left of the vertical line $r = r_*^{max}$ as seen in Fig.~\ref{fig:numgrids}.

It is important to note that the values of the finite constant $r_*^{max}$ are proportional to the value of the coupling $\ell_\eta$ i.e. $r_*^{max} \sim \ell_\eta$ (see Table~\ref{ref_table}). This means that the value of $\ell_\eta$ dictates the range of $r_*$ since $r_* \in (-\infty\,,\,r_*^{max}]$. A second important consequence of the above proportionality is that, as $\ell_\eta$ increases we also need to increase the number of grid points $N$ in order to keep the value of $\Delta r_*$ sufficiently small. To better understand why this is occurring we need to delve into the technical details concerning the procedure executed by our code.

The first step is to find the function $r(r_*)$ by numerically solving the differential equation of the tortoise coordinate
\begin{align}
	\frac{dr(r_*)}{dr_*} = \sqrt{\frac{f(r(r_*))}{g(r(r_*))}},
\end{align}
together with the condition $r(r_*=0)=1.00001\,r_{h}$ which fixes the integration constant. Hence, after the integration we have $r(r_*\rightarrow-\infty)\rightarrow r_{h}$, $r(r_*=0) = 1.00001\,r_{h}$ and $r(r_*\rightarrow r_*^{max})\rightarrow \infty$, meaning that $r_* \in (-\infty\,,\,r_*^{max}]$. However, in order to define a numerical grid with which we will perform the time evolution of $u$, we need to work on a finite interval of $r_*$. We do so by choosing a sufficiently large negative value\footnote{We note that this value is kept constant throughout all of our evolutions.} (which we denote by $r_*^{min}$) as the second end of the interval of $r_*$. Thus, in the context of the numerical integration we will work on the interval $r_*^{numerical} \in [r_*^{min}\,,\,r_*^{max}]$ even though in principle $r_* \in (-\infty\,,\,r_*^{max}]$.

The final step of our code which calculates the time domain profiles expects as inputs the values of $r_*^{min}\,,\,r_*^{max}$ and $N$ in the $r_*$ direction. It then calculates the spatial step $\Delta r_*$ from the relation
\begin{align}
	\Delta r_* = \frac{r_*^{max} + |r_*^{min}|}{N}\label{space_step_equation}
\end{align}
and the time step from $\Delta t = c\, \Delta r_*$ where $c$ is positive constant value satisfying the CFL condition that should not be confused with the speed of light. The fact that $r_*^{min}$ is constant throughout all of our evolutions and that $r_*^{max} \sim \ell_\eta$ implies, through Eq. \eqref{space_step_equation}, that as $\ell_
\eta$ increases we also need to increase $N$ in order to keep the value of $\Delta r_*$ sufficiently small (see Table~\ref{steps_table}).

\begin{table}[h]\centering
	\begin{tabular}{ |c|c|c| }
		\hline
		$\ell_\eta$ & $r_*^{max}$ \\[0.07cm]
		\hline
		0.1 & 1.017   \\
		5   & 29.501  \\
		100 & 351.929 \\
		\hline
	\end{tabular}
	\caption{Reference values indicating the analogy $r_*^{max} \sim \ell_\eta$.}
	\label{ref_table}
\end{table}

Finally, we can produce convergence curves to provide some quantitative information regarding the accuracy of our numerical integration scheme. To produce these curves, we first calculate the values of $u(r_*,t)$ at a given point as the grid spacing $\Delta r_*$ is reduced by increasing $N$. We will denote these values by $u(r_*,t)|_N$. We then use the value $u(r_*,t)$ for the maximum  number of grid points (i.e. the smaller grid spacing $\Delta r_*$) as a reference value indicating the best approximation to the true value of $u$ at that point. We will denote that value by $u(r_*,t)|_{best}$. To calculate the error we subtract each value $u(r_*,t)|_N$ for every different $N$ from the value of the best approximation and then take its absolute value i.e.
\begin{align}\label{error}
	|\text{Error}|_{N} = \Big| u(r_*,t)|_{best}-u(r_*,t)|_{N} \Big|~.
\end{align}

\begin{figure}[h]
	\centering
	\includegraphics[scale=0.55]{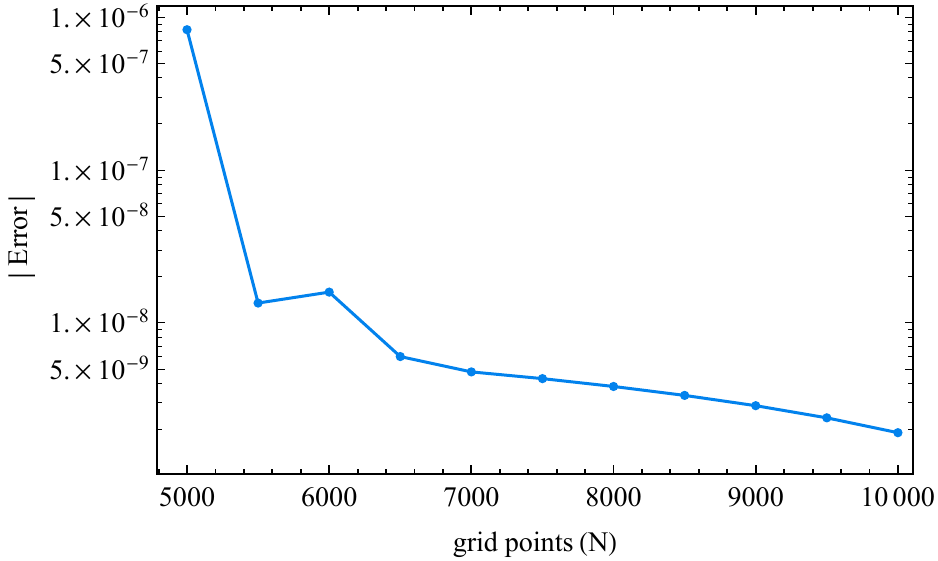}
	\includegraphics[scale=0.55]{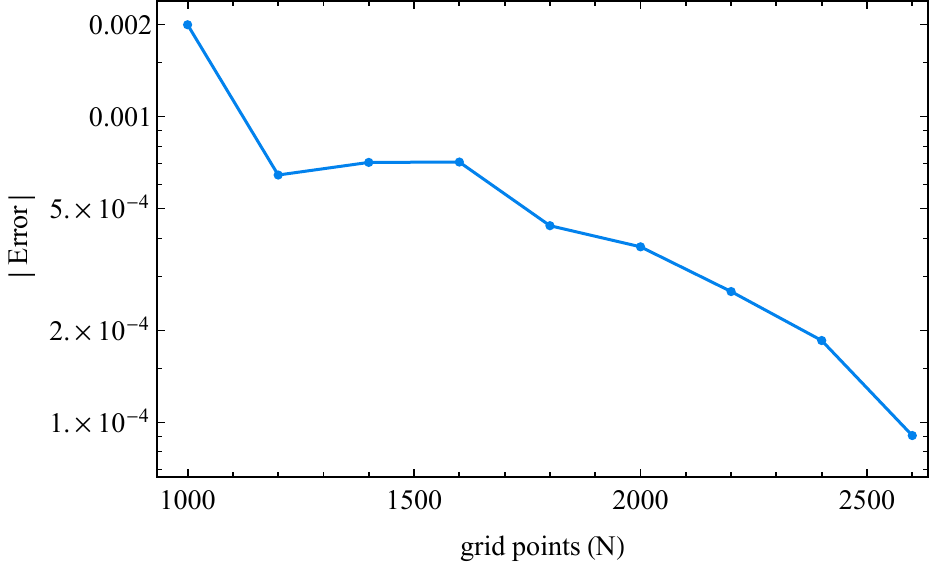}
	\caption{Left: Convergence curve for $m=0.1\,,\,\ell_\eta=100$ corresponding to a case where we obtain a clear prompt ringdown. As $u(r_*,t)|_{best}$ we choose the value of $u$ for $N=12000$ grid points, i.e. $u(r_*,t)|_{12000}$, indicating the best approximation. All the points are extracted at $r_*=0$ and $t/m=225.228$. Right: Convergence curve for $m=0.5\,,\,\ell_\eta=5$ corresponding to a case where we obtain echoes after the initial ringdown. As $u(r_*,t)|_{best}$ we choose the value of $u$ for $N=2800$ grid points, i.e. $u(r_*,t)|_{2800}$, indicating the best approximation. All the points are extracted at $r_*=0$ and $t/m=234.637$.}
	\label{fig:convergence_test}
\end{figure}

The diagrams in Fig.~\ref{fig:convergence_test} demonstrate that our code achieves numerical convergence irrespective of whether our compact object responds with a clear ringdown or with a signal with echoes i.e. in rather different regions of our parametric space $(m,\ell_\eta)$. Even though for the first case of Fig. \ref{fig:convergence_test} on the left the code converges rapidly, we expect that the same will occur for the second case depicted in Fig. \ref{fig:convergence_test} on the right if we further increase the number of grid points. As a final note, we stress the fact that even though the chosen values of the grid points $N$ are very different for the convergence curves in Fig.~\ref{fig:convergence_test}, the corresponding grid spacing $\Delta r_*$ is of the same order of magnitude for both cases, as can be seen in Table~\ref{steps_table}.

\begin{table}
	\begin{tabular}{ |c|c|c| }
		\hline
		\multicolumn{2}{|c|}{$\ell_\eta = 100\,,\, m = 0.1$} \\[0.07cm]
		\hline
		grid points ($N$) & $\Delta r_*$ \\[0.07cm]
		\hline
		5000 & 0.07  \\
		6000 & 0.058 \\
		7000 & 0.05  \\
		8000 & 0.043 \\
		9000 & 0.039 \\
		10000 & 0.035\\
		12000 & 0.029\\
		\hline
	\end{tabular}
	\begin{tabular}{ |c|c|c| }
		\hline
		\multicolumn{2}{|c|}{$\ell_\eta = 5\,,\, m = 0.5$} \\[0.07cm]
		\hline
		grid points ($N$) & $\Delta r_*$ \\[0.07cm]
		\hline
		1000 & 0.0695 \\
		1200 & 0.0579 \\
		1600 & 0.0434 \\
		2000 & 0.0347 \\
		2200 & 0.0316 \\
		2600 & 0.0267 \\
		2800 & 0.0248 \\
		\hline
	\end{tabular}
	\caption{The corresponding grid spacing $\Delta r_*$ for various choices of grid points $N$. Different values of grid points correspond to similar values of $\Delta r_*$ due to the different choices $\ell_\eta$. }
	\label{steps_table}
\end{table}


\begin{thebibliography}{99}

\bibitem{Abbott:2016blz}
LIGO Scientific and Virgo Collaborations collaboration, B.~P. Abbott
et~al., Observation of Gravitational Waves from a Binary Black Hole
Merger,  Phys. Rev. Lett. 116 (2016) 061102.

\bibitem{Abbott:2016nmj}
VGW151226: Observation of Gravitational Waves from a 22-Solar-Mass
Binary Black Hole Coalescence,
Phys. Rev. Lett. 116 (2016) 241103.

\bibitem{Abbott:2017vtc}
{\scshape VIRGO, LIGO Scientific} collaboration, B.~P. Abbott et~al.,
GW170104: Observation of a 50-Solar-Mass Binary Black Hole Coalescence
at Redshift 0.2,
Phys. Rev. Lett.  118 (2017) 221101.

\bibitem{Abbott:2017oio}
Virgo, LIGO Scientific collaboration, B.~P. Abbott et~al.,
GW170814: A Three-Detector Observation of Gravitational Waves from a
Binary Black Hole Coalescence,
Phys. Rev. Lett. 119 (2017) 141101.

\bibitem{TheLIGOScientific:2017qsa}
Virgo, LIGO Scientific collaboration, B.~P. Abbott et~al.,
GW170817: Observation of Gravitational Waves from a Binary Neutron
Star Inspiral,
Phys. Rev. Lett. 119 (2017) 161101.

\bibitem{Vishveshwara:1970zz}
C.~V.~Vishveshwara,
Scattering of Gravitational Radiation by a Schwarzschild Black-hole,
Nature {\bf 227}, 936 (1970).

\bibitem{Kokkotas:1999bd}
K.~D.~Kokkotas and B.~G.~Schmidt,
Quasinormal modes of stars and black holes,
Living Rev.\ Rel.\  {\bf 2}, 2 (1999).

\bibitem{Berti:2009kk}
E.~Berti, V.~Cardoso and A.~O.~Starinets,
Quasinormal modes of black holes and black branes,
Class.\ Quant.\ Grav.\  {\bf 26}, 163001 (2009).

\bibitem{Konoplya:2011qq}
R.~A.~Konoplya and A.~Zhidenko,
Quasinormal modes of black holes: From astrophysics to string theory,
Rev.\ Mod.\ Phys.\  {\bf 83}, 793 (2011).

\bibitem{Mazur:2001fv}
P.~O.~Mazur and E.~Mottola,
Gravitational condensate stars: An alternative to black holes,
[arXiv:gr-qc/0109035 [gr-qc]].

\bibitem{Morris:1988tu}
M.~S.~Morris, K.~S.~Thorne and U.~Yurtsever,
Wormholes, Time Machines, and the Weak Energy Condition,
Phys. Rev. Lett. \textbf{61}, 1446-1449 (1988).

\bibitem{Damour:2007ap}
T.~Damour and S.~N.~Solodukhin,
Wormholes as black hole foils,
Phys. Rev. D \textbf{76}, 024016 (2007).

\bibitem{Holdom:2016nek}
B.~Holdom and J.~Ren,
Not quite a black hole,
Phys. Rev. D \textbf{95}, no.8, 084034 (2017).

\bibitem{Abedi:2016hgu}
J.~Abedi, H.~Dykaar and N.~Afshordi,
``Echoes from the Abyss: Tentative evidence for Planck-scale structure at black hole horizons,''
Phys. Rev. D \textbf{96}, no.8, 082004 (2017).

\bibitem{Abedi:2020ujo}
J.~Abedi, N.~Afshordi, N.~Oshita and Q.~Wang,
Quantum Black Holes in the Sky,
Universe \textbf{6}, no.3, 43 (2020).

\bibitem{Destounis:2020kss}
K.~Destounis, A.~G.~Suvorov and K.~D.~Kokkotas,
``Testing spacetime symmetry through gravitational waves from extreme-mass-ratio inspirals,''
Phys. Rev. D \textbf{102}, no.6, 064041 (2020).

\bibitem{Destounis:2021mqv}
K.~Destounis, A.~G.~Suvorov and K.~D.~Kokkotas,
``Gravitational-wave glitches in chaotic extreme-mass-ratio inspirals,''
Phys. Rev. Lett. \textbf{126}, no.14, 141102 (2021).

\bibitem{Destounis:2021rko}
K.~Destounis and K.~D.~Kokkotas,
``Gravitational-wave glitches: Resonant islands and frequency jumps in nonintegrable extreme-mass-ratio inspirals,''
Phys. Rev. D \textbf{104}, no.6, 064023 (2021).

\bibitem{Peng:2019cmm}
Y.~Peng,
``Scalarization of horizonless reflecting stars: neutral scalar fields non-minimally coupled to Maxwell fields,''
Phys. Lett. B \textbf{804}, 135372 (2020).

\bibitem{Barausse:2019pri}
E.~Barausse,
``Black holes in General Relativity and beyond,''
MDPI Proc. \textbf{17}, no.1, 1 (2019).

\bibitem{Cardoso:2019rvt}
V.~Cardoso and P.~Pani,
Testing the nature of dark compact objects: a status report,
Living Rev. Rel. \textbf{22}, no.1, 4 (2019).

\bibitem{Cardoso:2016rao}
V.~Cardoso, E.~Franzin and P.~Pani,
Is the gravitational-wave ringdown a probe of the event horizon?,
Phys.\ Rev.\ Lett.\  \textbf{116} (2016) no.17, 171101.

\bibitem{Cardoso:2016oxy}
V.~Cardoso, S.~Hopper, C.~F.~B.~Macedo, C.~Palenzuela and P.~Pani,
Gravitational-wave signatures of exotic compact objects and of quantum corrections at the horizon scale,
Phys.\ Rev.\ D \textbf{94} (2016) no.8, 084031.

\bibitem{Lasenby:2002mc}
A.~Lasenby, C.~Doran, J.~Pritchard, A.~Caceres and S.~Dolan,
``Bound states and decay times of fermions in a Schwarzschild black hole background,''
Phys. Rev. D \textbf{72}, 105014 (2005).

\bibitem{Dolan:2007mj}
S.~R.~Dolan,
``Instability of the massive Klein-Gordon field on the Kerr spacetime,''
Phys. Rev. D \textbf{76}, 084001 (2007).

\bibitem{Vieira:2021xqw}
H.~S.~Vieira and K.~D.~Kokkotas,
``Quasibound states of Schwarzschild acoustic black holes,''
Phys. Rev. D \textbf{104}, no.2, 024035 (2021).

\bibitem{Vieira:2021ozg}
H.~S.~Vieira, K.~Destounis and K.~D.~Kokkotas,
``Slowly-rotating curved acoustic black holes: Quasinormal modes, Hawking-Unruh radiation, and quasibound states,''
Phys. Rev. D \textbf{105}, no.4, 045015 (2022).

\bibitem{Cheung:2021bol}
M.~H.~Y.~Cheung, K.~Destounis, R.~P.~Macedo, E.~Berti and V.~Cardoso,
``Destabilizing the Fundamental Mode of Black Holes: The Elephant and the Flea,''
Phys. Rev. Lett. \textbf{128}, no.11, 111103 (2022).

\bibitem{Clifton:2011jh}
T.~Clifton, P.~G.~Ferreira, A.~Padilla and C.~Skordis,
Modified Gravity and Cosmology,
Phys. Rept. \textbf{513}, 1-189 (2012).

\bibitem{Horndeski:1974wa}
G.W. Horndeski, {Second-order scalar-tensor field equations in a four-dimensional space}. Int. J. Theor. Phys. \textbf{10}, 363 (1974).

\bibitem{Nicolis:2008in}
A.~Nicolis, R.~Rattazzi, E.~Trincherini, {The Galileon as a local modification
of gravity}. Phys. Rev. D \textbf{79}, 064036 (2009).

\bibitem{Deffayet:2009wt}
C.~Deffayet, G.~Esposito-Farese, A.~Vikman, {Covariant Galileon}. Phys. Rev. D
\textbf{79}, 084003 (2009).

\bibitem{Ostrogradsky:1850fid}
M.~Ostrogradsky, {M\'{e}moires sur les \'{e}quations diff\'{e}rentielles, relatives au probl\`{e}me des isop\'{e}rim\`{e}tres}. Mem. Acad. St. Petersbourg \textbf{6}(4), 385
(1850).

\bibitem{Woodard:2006nt}
R.~P.~Woodard,
Avoiding dark energy with 1/r modifications of gravity,
Lect. Notes Phys. \textbf{720}, 403-433 (2007).

\bibitem{Woodard:2015zca}
R.~P.~Woodard,
Ostrogradsky's theorem on Hamiltonian instability,
Scholarpedia \textbf{10}, no.8, 32243 (2015).

\bibitem{Deffayet:2011gz}
C.~Deffayet, X.~Gao, D.~A.~Steer and G.~Zahariade,
From k-essence to generalised Galileons,
Phys. Rev. D \textbf{84}, 064039 (2011).

\bibitem{Amendola:1993uh}
L.~Amendola,
Cosmology with nonminimal derivative couplings,
Phys.\ Lett.\  B {\bf 301}, 175 (1993).

\bibitem{Sushkov:2009hk}
S.~V.~Sushkov,
Exact cosmological solutions with nonminimal derivative coupling,
Phys.\ Rev.\  D {\bf 80}, 103505 (2009).

\bibitem{Germani:2010hd}
C.~Germani, A.~Kehagias, {UV-Protected Inflation}. Phys. Rev. Lett.
\textbf{106}, 161302 (2011)

\bibitem{Koutsoumbas:2017fxp}
G.~Koutsoumbas, K.~Ntrekis, E.~Papantonopoulos and E.~N.~Saridakis,
``Unification of Dark Matter - Dark Energy in Generalized Galileon Theories,''
JCAP \textbf{02}, 003 (2018).

\bibitem{Dalianis:2019vit}
I.~Dalianis, S.~Karydas and E.~Papantonopoulos,
Generalized Non-Minimal Derivative Coupling: Application to Inflation and Primordial Black Hole Production,
JCAP \textbf{06}, 040 (2020).

\bibitem{Karydas:2021wmx}
S.~Karydas, E.~Papantonopoulos and E.~N.~Saridakis,
``Successful Higgs inflation from combined nonminimal and derivative couplings,''
Phys. Rev. D \textbf{104}, no.2, 023530 (2021).

\bibitem{DeFelice:2011bh}
A.~De Felice and S.~Tsujikawa,
``Conditions for the cosmological viability of the most general scalar-tensor theories and their applications to extended Galileon dark energy models,''
JCAP \textbf{02}, 007 (2012).

\bibitem{Papantonopoulos:2019eff}
E.~Papantonopoulos,
``Effects of the kinetic coupling of matter to curvature,''
Int. J. Mod. Phys. D \textbf{28}, no.05, 1942007 (2019).

\bibitem{Baldi:2005gk}
M.~Baldi, F.~Finelli and S.~Matarrese,
Inflation with violation of the null energy condition,
Phys. Rev. D \textbf{72}, 083504 (2005).

\bibitem{Libanov:2007mq}
M.~Libanov, V.~Rubakov, E.~Papantonopoulos, M.~Sami and S.~Tsujikawa,
UV stable, Lorentz-violating dark energy with transient phantom era,
JCAP \textbf{08}, 010 (2007).

\bibitem{Saridakis:2010mf}
E.~N.~Saridakis and S.~V.~Sushkov,
Quintessence and phantom cosmology with non-minimal derivative coupling,
Phys. Rev. D \textbf{81}, 083510 (2010).

\bibitem{Kolyvaris:2011fk}
T.~Kolyvaris, G.~Koutsoumbas, E.~Papantonopoulos, G.~Siopsis, {Scalar Hair from
a Derivative Coupling of a Scalar Field to the Einstein Tensor}. Class.
Quant. Grav. \textbf{29}, 205011 (2012)

\bibitem{Rinaldi:2012vy}
M.~Rinaldi,
Black holes with non-minimal derivative coupling,
Phys.\ Rev.\ D {\bf 86}, 084048 (2012).

\bibitem{Kolyvaris:2013zfa}
T.~Kolyvaris, G.~Koutsoumbas, E.~Papantonopoulos, G.~Siopsis, {Phase Transition
to a Hairy Black Hole in Asymptotically Flat Spacetime}. JHEP \textbf{11},
133 (2013)

\bibitem{Charmousis:2014zaa}
C.~Charmousis, T.~Kolyvaris, E.~Papantonopoulos, M.~Tsoukalas, {Black Holes in
Bi-scalar Extensions of Horndeski Theories}. JHEP \textbf{07}, 085 (2014).

\bibitem{Babichev:2013cya}
E.~Babichev and C.~Charmousis,
Dressing a black hole with a time-dependent Galileon,
JHEP \textbf{08} (2014), 106.

\bibitem{Bekenstein:1995un}
J.~D.~Bekenstein,
Novel \textquoteleft{}\textquoteleft{}no-scalar-hair\textquoteright{}\textquoteright{} theorem for black holes,
Phys. Rev. D \textbf{51}, no.12, R6608 (1995).

\bibitem{Hui:2012qt}
L.~Hui and A.~Nicolis,
No-Hair Theorem for the Galileon,
Phys. Rev. Lett. \textbf{110}, 241104 (2013).

\bibitem{Gubser:2008px}
S.~S.~Gubser,
Breaking an Abelian gauge symmetry near a black hole horizon,
Phys. Rev. D \textbf{78}, 065034 (2008).

\bibitem{Gubser:2005ih}
S.~S.~Gubser,
Phase transitions near black hole horizons,
Class. Quant. Grav. \textbf{22}, 5121-5144 (2005).

\bibitem{Hartnoll:2008vx}
S.~A.~Hartnoll, C.~P.~Herzog and G.~T.~Horowitz,
Building a Holographic Superconductor,
Phys. Rev. Lett. \textbf{101}, 031601 (2008).

\bibitem{Hartnoll:2008kx}
S.~A.~Hartnoll, C.~P.~Herzog and G.~T.~Horowitz,
Holographic Superconductors,
JHEP \textbf{12}, 015 (2008).

\bibitem{Kuang:2016edj}
X.~M.~Kuang and E.~Papantonopoulos,
``Building a Holographic Superconductor with a Scalar Field Coupled Kinematically to Einstein Tensor,''
JHEP \textbf{08}, 161 (2016).

\bibitem{Bea:2020ees}
Y.~Bea, O.~J.~C.~Dias, T.~Giannakopoulos, D.~Mateos, M.~Sanchez-Garitaonandia, J.~E.~Santos and M.~Zilhao,
``Crossing a large-$N$ phase transition at finite volume,''
JHEP \textbf{02}, 061 (2021).

\bibitem{Bea:2021zsu}
Y.~Bea, J.~Casalderrey-Solana, T.~Giannakopoulos, D.~Mateos, M.~Sanchez-Garitaonandia and M.~Zilh\~ao,
``Bubble wall velocity from holography,''
Phys. Rev. D \textbf{104}, no.12, L121903 (2021).

\bibitem{Bea:2021ieq}
Y.~Bea, J.~Casalderrey-Solana, T.~Giannakopoulos, D.~Mateos, M.~Sanchez-Garitaonandia and M.~Zilh\~ao,
``Domain Collisions,''
[arXiv:2111.03355 [hep-th]].

\bibitem{Bea:2021zol}
Y.~Bea, J.~Casalderrey-Solana, T.~Giannakopoulos, A.~Jansen, S.~Krippendorf, D.~Mateos, M.~Sanchez-Garitaonandia and M.~Zilh\~ao,
``Spinodal Gravitational Waves,''
[arXiv:2112.15478 [hep-th]].

\bibitem{Bea:2022mfb}
Y.~Bea, J.~Casalderrey-Solana, T.~Giannakopoulos, A.~Jansen, D.~Mateos, M.~Sanchez-Garitaonandia and M.~Zilh\~ao,
``Holographic Bubbles with Jecco: Expanding, Collapsing and Critical,''
[arXiv:2202.10503 [hep-th]].

\bibitem{Vlachos:2021weq}
C.~Vlachos, E.~Papantonopoulos and K.~Destounis,
Echoes of Compact Objects in Scalar-Tensor Theories of Gravity,
Phys. Rev. D \textbf{103}, no.4, 044042 (2021).

\bibitem{Mark:2017dnq}
Z.~Mark, A.~Zimmerman, S.~M.~Du and Y.~Chen,
A recipe for echoes from exotic compact objects,
Phys. Rev. D \textbf{96}, no.8, 084002 (2017).

\bibitem{Maselli:2017tfq}
A.~Maselli, S.~H.~V\"olkel and K.~D.~Kokkotas, Parameter estimation of gravitational wave echoes from exotic compact objects,
Phys. Rev. D \textbf{96}, no.6, 064045 (2017).

\bibitem{Volkel:2018hwb}
S.~H.~V\"olkel and K.~D.~Kokkotas,
Wormhole Potentials and Throats from Quasi-Normal Modes,
Class. Quant. Grav. \textbf{35}, no.10, 105018 (2018).

\bibitem{Konoplya:2018yrp}
R.~A.~Konoplya, Z.~Stuchl\'\i{}k and A.~Zhidenko,
Echoes of compact objects: new physics near the surface and matter at a distance,
Phys. Rev. D \textbf{99}, no.2, 024007 (2019).

\bibitem{Maggio:2019zyv}
E.~Maggio, A.~Testa, S.~Bhagwat and P.~Pani,
Analytical model for gravitational-wave echoes from spinning remnants,
Phys. Rev. D \textbf{100}, no.6, 064056 (2019).

\bibitem{Liu:2020qia}
H.~Liu, P.~Liu, Y.~Liu, B.~Wang and J.~P.~Wu,
Echoes from phantom wormholes,
Phys. Rev. D \textbf{103}, no.2, 024006 (2021).

\bibitem{Baker:2017hug}
T.~Baker, E.~Bellini, P.~G.~Ferreira, M.~Lagos, J.~Noller and I.~Sawicki,
``Strong constraints on cosmological gravity from GW170817 and GRB 170817A,''
Phys. Rev. Lett. \textbf{119}, no.25, 251301 (2017).

\bibitem{Creminelli:2017sry}
P.~Creminelli and F.~Vernizzi,
``Dark Energy after GW170817 and GRB170817A,''
Phys. Rev. Lett. \textbf{119}, no.25, 251302 (2017).

\bibitem{Gong:2017kim}
Y.~Gong, E.~Papantonopoulos and Z.~Yi,
``Constraints on scalar\textendash{}tensor theory of gravity by the recent observational results on gravitational waves,''
Eur. Phys. J. C \textbf{78}, no.9, 738 (2018).

\bibitem{Bahamonde:2019shr}
S.~Bahamonde, K.~F.~Dialektopoulos and J.~Levi Said,
``Can Horndeski Theory be recast using Teleparallel Gravity?,''
Phys. Rev. D \textbf{100}, no.6, 064018 (2019).

\bibitem{Bahamonde:2019ipm}
S.~Bahamonde, K.~F.~Dialektopoulos, V.~Gakis and J.~Levi Said,
``Reviving Horndeski theory using teleparallel gravity after GW170817,''
Phys. Rev. D \textbf{101}, no.8, 084060 (2020).

\bibitem{Bahamonde:2021dqn}
S.~Bahamonde, M.~Caruana, K.~F.~Dialektopoulos, V.~Gakis, M.~Hohmann, J.~Levi Said, E.~N.~Saridakis and J.~Sultana,
``Gravitational-wave propagation and polarizations in the teleparallel analog of Horndeski gravity,''
Phys. Rev. D \textbf{104}, no.8, 084082 (2021).

\bibitem{Chatzifotis:2020oqr}
N.~Chatzifotis, G.~Koutsoumbas and E.~Papantonopoulos,
``Formation of bound states of scalar fields in AdS-asymptotic wormholes,''
Phys. Rev. D \textbf{104}, no.2, 024039 (2021).

\bibitem{Korolev:2014hwa}
R.~V.~Korolev and S.~V.~Sushkov,
Exact wormhole solutions with nonminimal kinetic coupling,
Phys. Rev. D \textbf{90}, 124025 (2014).

\bibitem{Evseev:2017jek}
O.~A.~Evseev and O.~I.~Melichev,
No static spherically symmetric wormholes in Horndeski theory,
Phys. Rev. D \textbf{97} (2018) no.12, 124040.

\bibitem{Chandrasekhar}
S. Chandrasekhar, Mathematical theory of black holes
(Clarendon Press, Oxford, 2006).


\bibitem{Cardoso:2008bp}
V.~Cardoso, A.~S.~Miranda, E.~Berti, H.~Witek and V.~T.~Zanchin,
Geodesic stability, Lyapunov exponents and quasinormal modes,
Phys. Rev. D \textbf{79}, 064016 (2009).

\bibitem{Cardoso:2017soq}
V.~Cardoso, J.~L.~Costa, K.~Destounis, P.~Hintz and A.~Jansen,
Quasinormal modes and Strong Cosmic Censorship,
Phys. Rev. Lett. \textbf{120}, no.3, 031103 (2018).

\bibitem{Cardoso:2018nvb}
V.~Cardoso, J.~L.~Costa, K.~Destounis, P.~Hintz and A.~Jansen,
Strong cosmic censorship in charged black-hole spacetimes: still subtle,
Phys. Rev. D \textbf{98}, no.10, 104007 (2018).

\bibitem{Destounis:2018qnb}
K.~Destounis,
Charged Fermions and Strong Cosmic Censorship,
Phys. Lett. B \textbf{795}, 211-219 (2019).

\bibitem{Liu:2019lon}
H.~Liu, Z.~Tang, K.~Destounis, B.~Wang, E.~Papantonopoulos and H.~Zhang,
Strong Cosmic Censorship in higher-dimensional Reissner-Nordstr\"om-de Sitter spacetime,
JHEP \textbf{03}, 187 (2019).

\bibitem{Destounis:2019omd}
K.~Destounis, R.~D.~B.~Fontana, F.~C.~Mena and E.~Papantonopoulos,
Strong Cosmic Censorship in Horndeski Theory,
JHEP \textbf{10}, 280 (2019).

\bibitem{Destounis:2020pjk}
K.~Destounis, R.~D.~B.~Fontana and F.~C.~Mena,
Accelerating black holes: quasinormal modes and late-time tails,
Phys. Rev. D \textbf{102}, no.4, 044005 (2020).

\bibitem{Destounis:2020yav}
K.~Destounis, R.~D.~B.~Fontana and F.~C.~Mena,
Stability of the Cauchy horizon in accelerating black-hole spacetimes,
Phys. Rev. D \textbf{102}, no.10, 104037 (2020).

\bibitem{Bronnikov:2012ch}
K.~A.~Bronnikov, R.~A.~Konoplya and A.~Zhidenko,
Instabilities of wormholes and regular black holes supported by a phantom scalar field,
Phys. Rev. D \textbf{86}, 024028 (2012).

\bibitem{Abdalla:2018ggo}
E.~Abdalla, B.~Cuadros-Melgar, J.~de Oliveira, A.~B.~Pavan and C.~E.~Pellicer,
Vectorial and spinorial perturbations in Galileon Black Holes: Quasinormal modes, quasiresonant modes and stability,
Phys. Rev. D \textbf{99}, no.4, 044023 (2019).

\bibitem{Gundlach:1993tp}
C.~Gundlach, R.~H.~Price and J.~Pullin,
Late time behavior of stellar collapse and explosions: 1. Linearized perturbations,
Phys. Rev. D \textbf{49} (1994), 883-889.

\bibitem{Cardoso:2001bb}
V.~Cardoso and J.~P.~S.~Lemos,
Quasinormal modes of Schwarzschild anti-de Sitter black holes: Electromagnetic and gravitational perturbations,
Phys. Rev. D \textbf{64}, 084017 (2001).

\bibitem{Horowitz:1999jd}
G.~T.~Horowitz and V.~E.~Hubeny,
Quasinormal modes of AdS black holes and the approach to thermal equilibrium,
Phys. Rev. D \textbf{62}, 024027 (2000).

\bibitem{Berti:2007dg}
E.~Berti, V.~Cardoso, J.~A.~Gonzalez and U.~Sperhake,
``Mining information from binary black hole mergers: A Comparison of estimation methods for complex exponentials in noise,''
Phys. Rev. D \textbf{75}, 124017 (2007).

\bibitem{Konoplya:2017wot}
R.~A.~Konoplya and Z.~Stuchl\'\i{}k,
``Are eikonal quasinormal modes linked to the unstable circular null geodesics?,''
Phys. Lett. B \textbf{771}, 597-602 (2017).

\bibitem{Holzegel:2011uu}
G.~Holzegel and J.~Smulevici,
``Decay properties of Klein-Gordon fields on Kerr-AdS spacetimes,''
Commun. Pure Appl. Math. \textbf{66}, 1751-1802 (2013).

\bibitem{Churilova:2019cyt}
M.~S.~Churilova and Z.~Stuchlik,
Ringing of the regular black-hole/wormhole transition,
Class. Quant. Grav. \textbf{37}, no.7, 075014 (2020).

\bibitem{Torii:1996yi}
T.~Torii, H.~Yajima and K.~i.~Maeda,
``Dilatonic black holes with Gauss-Bonnet term,''
Phys. Rev. D \textbf{55}, 739-753 (1997).

\bibitem{Blazquez-Salcedo:2018jnn}
J.~L.~Bl\'azquez-Salcedo, D.~D.~Doneva, J.~Kunz and S.~S.~Yazadjiev,
``Radial perturbations of the scalarized Einstein-Gauss-Bonnet black holes,''
Phys. Rev. D \textbf{98}, no.8, 084011 (2018).

\bibitem{Zou:2020zxq}
D.~C.~Zou and Y.~S.~Myung,
``Radial perturbations of the scalarized black holes in Einstein-Maxwell-conformally coupled scalar theory,''
Phys. Rev. D \textbf{102}, no.6, 064011 (2020).

\bibitem{Blazquez-Salcedo:2022omw}
J.~L.~Bl\'azquez-Salcedo, D.~D.~Doneva, J.~Kunz and S.~S.~Yazadjiev,
``Radial perturbations of scalar-Gauss-Bonnet black holes beyond spontaneous scalarization,''
[arXiv:2203.00709 [gr-qc]].

\bibitem{Minamitsuji:2014hha}
M.~Minamitsuji,
``Black hole quasinormal modes in a scalar-tensor theory with field derivative coupling to the Einstein tensor,''
Gen. Rel. Grav. \textbf{46}, 1785 (2014).

\bibitem{Dong:2017toi}
R.~Dong, J.~Sakstein and D.~Stojkovic,
``Quasinormal modes of black holes in scalar-tensor theories with nonminimal derivative couplings,''
Phys. Rev. D \textbf{96}, no.6, 064048 (2017).

\bibitem{Cardoso:2021wlq}
V.~Cardoso, K.~Destounis, F.~Duque, R.~P.~Macedo and A.~Maselli,
``Black holes in galaxies: Environmental impact on gravitational-wave generation and propagation,''
Phys. Rev. D \textbf{105}, no.6, L061501 (2022).

\bibitem{Ezquiaga:2022nak}
J.~M.~Ezquiaga, W.~Hu, M.~Lagos, M.~X.~Lin and F.~Xu,
``Modified gravitational wave propagation with higher modes and its degeneracies with lensing,''
[arXiv:2203.13252 [gr-qc]].

\bibitem{Deffayet:2021nnt}
C.~Deffayet, S.~Mukohyama and A.~Vikman,
``Ghosts without runaway,''
[arXiv:2108.06294 [gr-qc]].

\end{thebibliography}
\end{document}